\documentclass[useAMS,usenatbib,usegraphicx]{mn2e}
\usepackage{times}
\usepackage{amsmath}
\usepackage{amssymb}

\input{psfig.sty}

\newcommand{\AIPS}{{$\cal AIPS\/$}}

\def\qfir{$q_{\rm IR}$}
\def\qtwofifty{$q_{\rm 250}$}
\def\lfir{$L_{\rm IR}$}
\def\sfir{$S_{\rm IR}$}

\def\td{$T_{\rm d}$}
\def\gs{\mathrel{\raise0.35ex\hbox{$\scriptstyle >$}\kern-0.6em
\lower0.40ex\hbox{{$\scriptstyle \sim$}}}}
\def\ls{\mathrel{\raise0.35ex\hbox{$\scriptstyle <$}\kern-0.6em
\lower0.40ex\hbox{{$\scriptstyle \sim$}}}}
\def\m@th{\mathsurround=0pt }
\def\eqalign#1{\null\,\vcenter{\openup1\jot \m@th
 \ialign{\strut\hfil$\displaystyle{##}$&$\displaystyle{{}##}$\hfil
 \crcr#1\crcr}}\,}

\title[Bolometric FIR/radio correlation]
      {BLAST: the far-infrared/radio correlation
       in distant galaxies}
\author[Ivison et al.]
       {R.\,J.\ Ivison,$^{\! 1,2}$
David M.\ Alexander,$^{\! 3}$ 
Andy D.\ Biggs,$^{\! 4}$
W.\,N.\ Brandt,$^{\! 5}$
Edward L.\ Chapin,$^{\! 6}$  \and
Kristen E.\,K.\ Coppin,$^{\! 3}$ 
Mark J.\ Devlin,$^{\! 7}$
Mark Dickinson,$^{\! 8}$
James Dunlop,$^{\! 2}$
Simon Dye,$^{\! 9}$ \and
Stephen A. Eales,$^{\! 9}$
David T.\ Frayer,$^{\! 10}$
Mark Halpern,$^{\! 6}$ 
David H.\ Hughes,$^{\! 11}$
Edo Ibar,$^{\! 1}$  \and
A.\ Kov\'{a}cs,$^{\! 12}$
Gaelen Marsden,$^{\! 6}$
L.\ Moncelsi,$^{\! 9}$ 
Calvin B.\ Netterfield,$^{\! 13,14}$ 
Enzo Pascale,$^{\! 9}$ \and
Guillaume Patanchon,$^{\! 15}$
D.\,A.\ Rafferty,$^{\! 5}$
Marie Rex,$^{\! 7}$ 
Eva Schinnerer,$^{\! 16}$
Douglas Scott,$^{\! 6}$ \and
C.\ Semisch,$^{\! 7}$
Ian Smail,$^{\! 3}$
A.\,M.\ Swinbank,$^{\! 3}$
Matthew D.\,P.\ Truch,$^{\! 7}$
Gregory S.\ Tucker,$^{\! 17}$ \and
Marco P.\ Viero,$^{\! 14}$ 
Fabian Walter,$^{\! 16}$ 
Axel Wei\ss,$^{\! 12}$
Donald V.\ Wiebe$^{6,14}$ and
Y.\,Q.\ Xue$^{5}$
 \vspace*{1mm}\\
$^1$ UK Astronomy Technology Centre, Royal Observatory, 
     Blackford Hill, Edinburgh EH9 3HJ\\
$^2$ Institute for Astronomy, University of Edinburgh, Blackford Hill,
     Edinburgh EH9 3HJ\\
$^3$ Institute for Computational Cosmology, Durham University,
     South Road, Durham DH1 3LE\\
$^4$ European Southern Observatory, Karl-Schwarzschild-Str.\ 2, D-85748, Germany\\
$^5$ Department of Astronomy and Astrophysics, The Pennsylvania State University,
     University Park, PA 16802, USA\\
$^6$ Department of Physics \& Astronomy, University of British Columbia,
     6224 Agricultural Road, Vancouver, BC V6T 1Z1, Canada\\
$^7$ Department of Physics \& Astronomy, University of Pennsylvania, 209
     South 33rd Street, Philadelphia, PA, 19104, USA\\
$^8$ NOAO, 950 N.\ Cherry Avenue, Tucson, AZ 85719, USA\\
$^9$ School of Physics and Astronomy, Cardiff University, Queens Buildings,
     The Parade, Cardiff CF24 3AA\\
$^{10}$ Spitzer Science Center, California Institute of Technology,
        Pasadena, CA 91125, USA\\
$^{11}$ Instituto Nacional de Astrof\'isica \'Optica y Electr\'onica, 
        Aptdo.\ Postal 51 y 72000 Puebla, Mexico\\
$^{12}$ Max-Planck Institute f\"ur Radioastronomie, D-53121 Bonn, Germany\\
$^{13}$ Department of Astronomy \& Astrophysics, University of Toronto, 50
        St.\ George Street Toronto, ON M5S~3H4, Canada\\
$^{14}$ Department of Physics, University of Toronto, 60 St.\ George Street,
        Toronto, ON M5S~1A7, Canada\\
$^{15}$ Laboratoire APC, 10, rue Alice Domon et L\'eonie Duquet,
        75205 Paris, France\\
$^{16}$ Max-Planck-Institut f\"ur Astronomie, K\"onigstuhl 17, D-69117
        Heidelberg, Germany\\
$^{17}$ Department of Physics, Brown University, 182 Hope Street, Providence,
        RI 02912, USA}

\date{Accepted ... ; Received ... ; in original form ...}

\pagerange{000--000}

\begin{document}

\maketitle

\begin{abstract}
  We investigate the correlation between far-infrared (FIR) and radio
  luminosities in distant galaxies, a lynchpin of modern astronomy. We
  use data from the Balloon-borne Large Aperture Submillimetre
  Telescope (BLAST), {\em Spitzer}, the Large Apex BOlometer CamerA
  (LABOCA), the Very Large Array (VLA) and the Giant Metre-wave Radio
  Telescope (GMRT) in the Extended {\em Chandra} Deep Field South
  (ECDFS). For a catalogue of BLAST 250-$\mu$m-selected galaxies, we
  re-measure the 70--870-$\mu$m flux densities at the positions of
  their most likely 24-$\mu$m counterparts, which have a median
  [interquartile] redshift of 0.74 [0.25, 1.57]. From these, we
  determine the monochromatic flux density ratio, \qtwofifty\ (=
  log$_{10}$ [$S_{\rm 250\mu m}/S_{\rm 1,400MHz}$]), and the
  bolometric equivalent, \qfir.  At $z\approx 0.6$, where our
  250-$\mu$m filter probes rest-frame 160-$\mu$m emission, we find no
  evolution relative to $q_{\rm 160}$ for local galaxies. We also
  stack the FIR and submm images at the positions of 24-$\mu$m- and
  radio-selected galaxies. The difference between \qfir\ seen for
  250-$\mu$m- and radio-selected galaxies suggests star formation
  provides most of the IR luminosity in $\ls$100-$\mu$Jy radio
  galaxies, but rather less for those in the mJy regime. For the
  24-$\mu$m sample, the radio spectral index is constant across
  $0<z<3$, but \qfir\ exhibits tentative evidence of a steady decline
  such that \qfir\ $\propto (1+z)^{-0.15\pm0.03}$ -- significant
  evolution, spanning the epoch of galaxy formation, with major
  implications for techniques that rely on the FIR/radio
  correlation. We compare with model predictions and speculate that we
  may be seeing the increase in radio activity that gives rise to the
  radio background.
\end{abstract}


\section{Introduction}

The correlation between FIR and radio luminosities
\citep[e.g.][]{vanderkruit71, dickey84, dejong85, helou85} is believed
to be due to a common link with massive, dust-enshrouded stars. During
their brief lives these stars warm the dusty molecular clouds in which
they were born, expel dust in their winds, then create radio-luminous
supernova remnants (which are believed to accelerate cosmic-ray
electrons), perhaps generating more dust during this finale
\citep[e.g.][]{dunne09}.

The empirical FIR/radio correlation is regularly exploited in a
variety of ways -- to calibrate the relationship between radio
luminosity and star-formation rate \citep{condon92, bell03}, for
example, or to estimate the distance to luminous starbursts
\citep[e.g.][]{cy99}, or their dust temperatures \citep[$T_{\rm d}$ --
e.g.][]{chapman05}, or to define samples of radio-excess active
galactic nuclei \citep[AGN -- e.g.][]{donley05}. The correlation has
thus become a cornerstone of modern astronomy. It is clearly important
to know whether the correlation breaks down at extreme luminosities,
or varies with redshift, perhaps due to variations in magnetic field
strength \citep[e.g.][]{bernet08}, IR photon density, initial mass
function (IMF), dust composition or cosmic-ray flux
\citep[e.g.][]{rengarajan05}.

Recent efforts have concentrated on determining the relationship
between radio and 24-$\mu$m flux densities, $q_{24}$ = ${\rm log}_{10}
\, (S_{\rm 24\mu m}/S_{\rm 1,400MHz})$, taking advantage of {\em
Spitzer}'s sensitivity to hot dust emission from very distant
star-forming galaxies \citep[e.g.][]{appleton04, ibar08}. A galaxy's
24-$\mu$m luminosity is not a particularly reliable tracer of its FIR
luminosity, however, being prone to uncertain contamination by AGN
continuum emission and by spectral features due to silicates and
polycyclic aromatic hydrocarbons \citep[PAHs --][]{pope06,desai07}.

\citeauthor{appleton04} noted that `Ideally, it would be better to
measure a bolometric $q$, but insufficient data are available at
longer wavelengths to do this reliably'. Here, we exploit new data
from BLAST \citep{devlin09} and the LABOCA ECDFS Submm Survey
\citep[LESS --][]{weiss09} to explore the correlation between {\it
bolometric} IR luminosity and $K$-corrected radio luminosity. We do
this separately for FIR-, 24-$\mu$m- and radio-selected galaxies. The
latter catagories are explored in \S\ref{stacking}, stacking at the
positions of 24-$\mu$m and radio emitters.

Before we describe our analysis, the question of how best to explore
the FIR/radio correlation for FIR-selected galaxies merits discussion.
A thorough treatment of the IR and radio properties of FIR-selected
galaxies must be able to deal with the potentially FIR-loud and
radio-weak fractions of the sample. If the available FIR data were
well matched in resolution to typical optical/IR or radio imaging then
comparing FIR and radio properties would be trivial. As things stand,
however, the BLAST 250-$\mu$m beam covers $\sim$500$\times$ more area
than a typical VLA synthesised beam and matching FIR and radio sources
is non-trivial (\S\ref{stats}).

We must consider how to deal with cases where counterparts to our FIR
emitters cannot be found. This leaves us with no way to determine
their positions unambigiously, or even their FIR flux densities since
they are often blended, or their radio flux densities (or appropriate
limits), or their redshifts. We can not begin the analysis by
pinpointing the FIR emitters in the radio waveband since this
introduces a strong bias in favour of the most luminous radio
emitters. \citet{marsden09} showed that the 24-$\mu$m galaxy
population detected by {\em Spitzer} \citep{magnelli09} can account
for all of the FIR background, as measured by the {\em Cosmic
  Background Explorer} \citep{puget96, fixsen98}. We have already
noted that the correlation between 24-$\mu$m and FIR luminosities is
poor, but every FIR source should have a 24-$\mu$m counterpart and our
approach to pinning down the positions and radio properties of the
parent sample is to adopt the 24-$\mu$m identifications proposed by
\citet{dye09} or, where \citeauthor{dye09} opts for a radio
identification, our own 24-$\mu$m identifications. Armed with these
positions, we determine the radio properties and redshifts of the
appropriate counterparts, including appropriate limits for any that
lack radio emission.

Our paper is laid out as follows. In \S\ref{observations} we introduce
a 250-$\mu$m-selected sample of galaxies from BLAST and the {\em
Spitzer}, LABOCA, VLA and GMRT data used to determine their IR and
radio spectral energy distributions (SEDs). In \S\ref{stats} we
cross-match our FIR sample with radio emitters, demonstrating how
difficult this will be for deep surveys with {\em Herschel}. We
cross-match with the most likely 24-$\mu$m counterparts
(\S\ref{24um}), enabling us to explore the FIR/radio correlation for
our 250-$\mu$m-selected galaxies in \S\ref{results}, unbiased by radio
selection. We also investigate the correlation for 24-$\mu$m- and
radio-selected samples (\S\ref{stacking} and \ref{radstack}), via the
stacking technique \citep[e.g.][]{ivison07b,dunne09b,marsden09},
looking at the effect of $K$ corrections using several SED
templates. We state our conclusions in \S\ref{conclusions}.

Throughout this paper we assume a Universe with $\Omega_m=0.27$,
$\Omega_\Lambda=0.73$ and $H_0=71$\,km\,s$^{-1}$\,Mpc$^{-1}$
\citep{spergel07}.

\section{Observations}
\label{observations}

Our primary dataset, known as `BLAST GOODS South -- Deep' \citep[where
GOODS is the Great Observatories Origins Deep Survey
--][]{dickinson03} was taken during BLAST's Antarctic flight in 2006
\citep{devlin09} and comprises maps\footnote{Available at {\tt
http://blastexperiment.info}} covering the ECDFS at 250, 350 and
500\,$\mu$m. We combine them with panchromatic data, including the
deepest available 70--160-$\mu$m imaging from {\em Spitzer} -- from
the Far-Infrared Deep Extragalactic Legacy survey (FIDEL), with
870-$\mu$m LABOCA imaging from the 12-m Atacama Pathfinder EXperiment
(APEX) telescope \citep{weiss09}, with high-resolution 1,400-MHz
imaging from the VLA \citep{miller08, biggs09} and with new 610-MHz
imaging from the GMRT. This latter dataset allows us to $K$-correct
the 1,400-MHz radio luminosity accurately out to $z\sim 1.3$, and
beyond if the spectra are intrinsic power laws.  Previous work assumes
radio emitters share a common spectral index, $\alpha\approx-0.8$
(where $S\propto\nu^{\alpha}$) -- a poor approximation, as shown by
\citet{ibar09}. Together, these data are capable -- prior to
operations with {\em Herschel} -- of determining the bolometric IR
luminosities of distant galaxies.

\begin{figure}
\begin{center}
  \includegraphics[scale=0.40,angle=270]{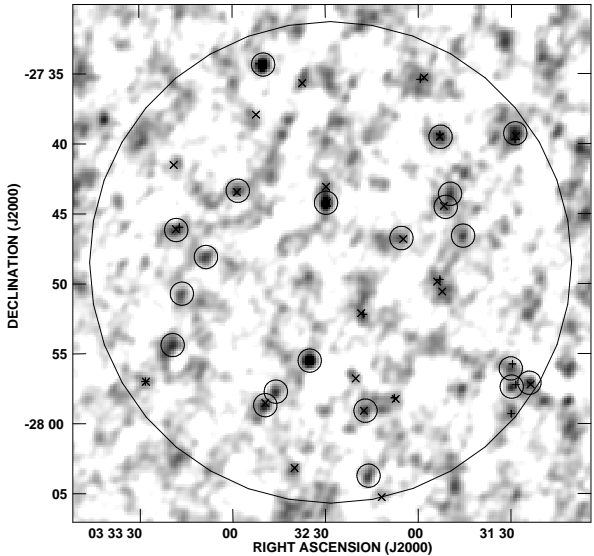}
  \caption{BLAST 250-$\mu$m image of the central region of ECDFS, with
    $>$3-$\sigma$ sources circled. $\sigma$ includes instrumental and
    confusion noise combined in quadrature, where $\sigma_{\rm
      confusion}/\sigma_{\rm instrumental}\sim 2$
    \citep{marsden09}. For the deepest imaging planned for {\em
      Herschel}, this ratio will be $\approx$25. Similarly significant
    sources at 350 and 500\,$\mu$m are labelled with $\times$ and $+$
    symbols, respectively. The region covered by the deepest VLA
    imaging, with relatively low levels of bandwidth smearing, is
    within the 17.2-arcmin-radius circle.}
  \label{blastmap}
\end{center}
\end{figure}

In order to take full advantage of the best available multi-wavelength
coverage, we limit the areal coverage of this study to that of the
deep VLA imaging described in \S\ref{vla}, a region of radius
17.2\,arcmin centred on $\alpha_{\rm J2000} = 03^{\rm h} 32^{\rm m}
28^{\rm s}.3, \delta_{\rm J2000} = -27^{\circ} 48' 30''$ (see
Fig.~\ref{blastmap}). This area coincides with the deepest of the
BLAST surveys.

\subsection{BLAST 250-, 350- and 500-$\mu$m measurements}
\label{blast}

The 1.8-m Balloon-borne Large Aperture Telescope, BLAST, is a
forerunner of the Spectral and Photometric Imaging Receiver
\citep[SPIRE --][]{griffin09} on the 3.5-m {\em Herschel Space
Observatory}.  Data here are from the successful 2006 flight, lasting
270\,hr, of which 90\,hr was spent surveying a field centred on the
ECDFS.

In the part of this survey used here, the mean instrumental
[confusion] noise levels are 11 [21], 9 [17] and 6
[15]\,mJy\,beam$^{-1}$ at 250, 350 and 500\,$\mu$m, respectively,
where the beams had {\sc fwhm} of 36, 42 and 60\,arcsec
\citep{devlin09, marsden09} and the noise values were determined in
beam-smoothed maps.

As our parent catalogue we adopt the sample of 250-$\mu$m BLAST
sources described by \citet{devlin09}, cutting at a signal-to-noise
ratio (SNR) of three, where the uncertainty in flux comprised the
quadrature sum of the instrumental noise (used to define SNR in the
\citeauthor{devlin09} catalogue) and the confusion noise. The
reliability of our sample is thus similar to those of the first submm
galaxy (SMG) catalogues \citep[e.g.][]{smail97, hughes98,
  barger98}. Thus defined, our parent catalogue comprises 22 sources,
listed in Table~\ref{squad} and marked in Fig.~\ref{blastmap}.

For the FIR measurements described in the sections that follow, the
BLAST image in each of the three filters has been convolved with the
appropriate noise-weighted point spread function \citep[PSF
--][]{truch09}. This is equivalent to calculating the
maximum-likelihood point-source flux density by which the PSF would
have been scaled to fit an isolated point source centred in each
pixel. Flux densities in each band are taken from these convolved
maps at the appropriate positions.

Since the differential 250-$\mu$m source counts fall rapidly with flux
\citep[${\rm d}N/{\rm d}S\propto S^{-3.5}$ -- see][]{patanchon09}, the
effects of Eddington bias and source confusion have conspired to boost
the fluxes in our parent catalogue. To determine the degree of
boosting, we simulate these effects by injecting a source population
consistent with the BLAST counts \citep{patanchon09} into a Gaussian
noise field resembling the instrumental fluctuations in the BLAST
GOODS South -- Deep field and then measuring the flux densities of
those sources at the input positions \citep[see Appendix B of
][]{eales09}. These simulations allow us to apply appropriate
first-order corrections and to estimate the uncertainties associated
with those corrections (see later in \S\ref{seds}). The corrections
are statistical in nature and can be tailored to individual sources
only in the sense that a source with a given SNR is likely to have
been boosted by factor, $B({\rm SNR})$, where $B$ ranges from
$\sim$1.1--1.6 for the range of SNRs present in our sample. A full
treatment of flux boosting in the BLAST data -- including the
influence of clustering and multi-band source selection -- will be
explored in future work.

\subsection{{\em Spitzer} 70- and 160-$\mu$m measurements}
\label{spitzer}

All of the raw FIDEL (P.I.: Dickinson), GO (P.I.: Frayer), and GTO
(P.I.: Rieke) {\em Spitzer} data covering ECDFS were reduced and
combined using the techniques developed by \citet{frayer06}. We
adopted the updated calibration factors, including corrections for
colour and for the point-source-response function (PRF) given by
\citet{frayer09}.

The {\em Spitzer} flux density measurements are made at the positions
discussed later and the quoted uncertainties represent the r.m.s.\
noise at those positions, combined with an additional 10 (15) per cent
uncertainty at 70\,$\mu$m (160\,$\mu$m) due to possible systematics
affecting the calibration factors.

\subsection{LABOCA 870-$\mu$m measurements}
\label{laboca}

LESS covers around 900\,arcmin$^2$ of ECDFS at a wavelength of
870\,$\mu$m, with a beamwidth of 19.2\,arcsec {\sc fwhm}
\citep[]{greve09, coppin09, weiss09}, to an average depth of
$\sigma=1.2$\,mJy\,beam$^{-1}$ as measured in a beam-smoothed map
across the region in which our 250-$\mu$m sample was selected. The
870-$\mu$m flux densities, at the positions discussed later, are
measured in the same manner discussed in \S\ref{blast} and
\ref{spitzer}.

\subsection{VLA 1,400-MHz imaging}
\label{vla}

Deep, high-resolution 1,400-MHz imaging of the ECDFS was described by
\citet{miller08} and we use their image to identify radio
counterparts for our parent 250-$\mu$m-selected source catalogue (see
\S\ref{ids}).

We use the techniques described by \citet{ibar09} to generate a new,
deeper 1,400-MHz catalogue and to correct for effects such as flux
boosting at low SNRs. This catalogue was cut at $>$4\,$\sigma$, where
$\sigma$ was determined from the local background.

In cases where the emission is found to be heavily resolved, e.g.\
for the large spiral associated with BLAST\,J033235$-$275530, we
measure the total radio flux density using {\sc tvstat} within \AIPS.

\subsection{GMRT 610-MHz imaging}
\label{gmrt}

\begin{figure*}
\begin{center}
  \includegraphics[scale=0.64]{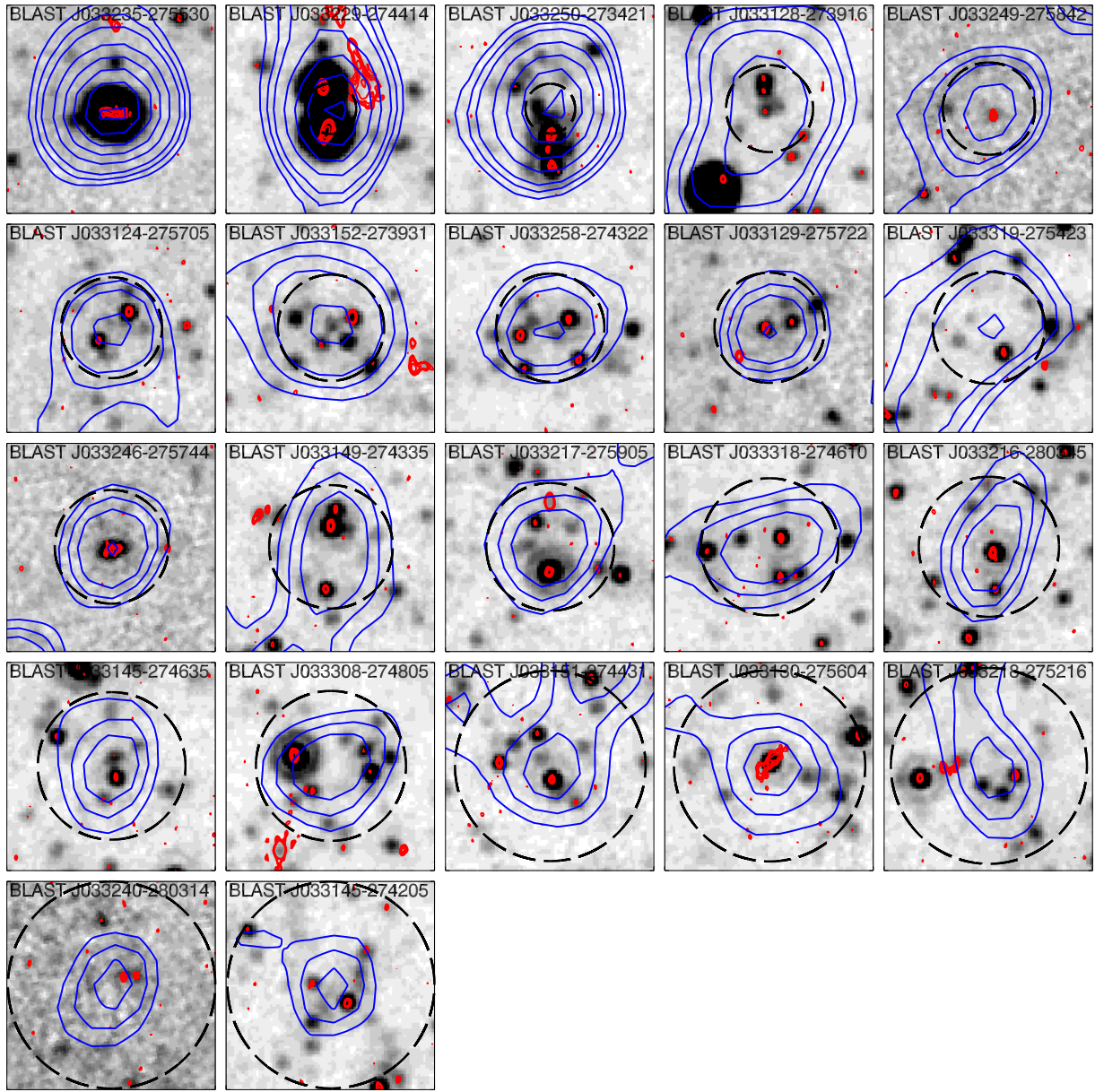}
  \caption{{\em Spitzer} 24-$\mu$m greyscale images superimposed with
    red 1,400-MHz contours and blue 250-$\mu$m contours, centred on
    BLAST 250-$\mu$m-selected sources which are labelled with their
    names. Each postage stamp image is
    90\,arcsec\,$\times$\,90\,arcsec. Dashed black circles show our
    search areas (3\,$\sigma_{\rm pos}$ radius -- see
    \S\ref{ids}). Broadly speaking, the sources are associated with
    either large spirals or clumps of faint, 24-$\mu$m galaxies --
    sometimes several of the former or as many as seven of the
    latter. The number of double sources is striking, especially those
    where the 250-$\mu$m emission peaks between the two.}
  \label{postage}
\end{center}
\end{figure*}

To test the feasibility of deep radio observations in ECDFS, a small
amount of new 610-MHz data were obtained using the GMRT\footnote{We
thank the staff of the GMRT who made these observations
possible. GMRT is run by the National Centre for Radio Astrophysics of
the Tata Institute of Fundamental Research.} in 2008 November. During
seven 9-hr sessions we obtained data centred on the six positions
chosen by \citet{miller08} for their VLA survey (\S\ref{vla}),
recording 128 channels every 16\,s in the lower and upper sidebands
(602 and 618\,MHz, respectively) in each of two polarisations. The
integration time in each field was $\sim$3\,hr, made up of 30-min
scans sandwiched between 5-min scans of the nearby calibrator,
0240$-$231, with 10-min scans of 3C\,48 and 3C\,147 for flux and
bandpass calibration.

Calibration followed standard recipes within \AIPS\ ({\sc 31dec09}). A
bandpass table was generated after self-calibrating 3C\,48 and 3C\,147
in phase with a solution interval of 1\,min. This was applied during
subsequent loops of calibration and flagging (with {\sc uvflg}, {\sc
tvflg}, {\sc spflg} and {\sc flgit}). The error-weighted flux
densities of 0240$-$231 were found to be $5.55\pm0.02$ and
$5.31\pm0.02$\,Jy at 602 and 618\,MHz, respectively; the r.m.s.\
scatter of the flux density measurements over the seven days was 0.12
and 0.10\,Jy which suggests that the flux calibration should be
accurate to better than 5 per cent.

The calibrated data were averaged down to yield 41 channels in each
sideband and the data from the seven days were concatenated using {\sc
dbcon}. The resulting datasets were then self-calibrated in phase
({\sc solmode=p!a}), with a solution interval of 2\,min, using {\sc
clean} components from our own reduction of the VLA A-configuation
data. This process yielded accurate phases; the astrometric reference
frame was also tied to that of the 1,400-MHz image as a result.
Subsequent imaging entailed the creation of a mosaic of 37 facets --
to cover most of the primary beam in each of the six pointings -- each
facet with 512$^2$\,pixels (1.5$^2$-arcsec$^2$ per pixel). A further
10--20 bright sources outside these central regions, identified in
heavily tapered maps, were also imaged for each pointing. Subsequent
self-calibration was carried out in phase alone, then in amplitude and
phase (first with {\sc cparm = 0, 1, 0} then with {\sc cparm = 0}),
with a solution interval of 2\,min, staggered by 1\,min. The $uv$ data
were weighted using {\sc robust} = $-$0.5, {\sc uvrange} = 0.9,
100\,k$\lambda$, {\sc guard} = $-$1, {\sc uvtaper} = 30,
60\,\,k$\lambda$ and {\sc uvbox} = 10.

After {\sc clean} components were subtracted from the $uv$ data, more
manual flagging was undertaken. Almost one third of the baselines were
rejected, in total. {\sc clean} components were re-introduced ({\sc
  uvsub, factor=}$-$1), then the final six mosaics were convolved to a
common beam size (6.5\,arcsec $\times$ 5.4\,arcsec, with the major
axis at position angle 174$^{\circ}$), then knitted together using
{\sc flatn}. An appropriate correction was made for the shape of the
primary beam, with data rejected at radii beyond the half-maximum
level. The final image has a noise level of
$\sim$40\,$\mu$Jy\,beam$^{-1}$ and covers the entire VLA image, with
insignificant levels of bandwidth smearing. We exploit it here to
determine the 610-MHz flux densities, and hence the spectral indices
between 610 and 1,400\,MHz, $\alpha^{\rm 1,400}_{\rm 610}$, of the
radio identifications described in \S\ref{ids}. These measurements are
listed in Table~\ref{squad}. Our 610-MHz image is available on
request\footnote{E-mail: rji@roe.ac.uk}.

\begin{table*}
\caption{Parent 250-$\mu$m catalogue from BLAST \citep{devlin09}, 24-$\mu$m identifications, radio and X-ray properties, and redshifts. }
\label{squad}
{\scriptsize
\begin{center}
\begin{tabular}{lcccccccccccc}
\hline
BLAST name&$\alpha_{\rm 250\mu m}$&$\delta_{\rm 250\mu m}$&$S_{\rm 250\mu m}^{\dagger}$&$\alpha_{\rm 24\mu m}$&$\delta_{\rm 24\mu m}$&$S_{\rm 24\mu m}$&$S_{\rm 1,400MHz}$&$S_{\rm 610Hz}$&$\alpha^{\rm 1,400}_{\rm 610}$&X-ray?$^{\ddagger}$&Spec&Phot\\
          & J2000     & J2000       &/mJy             & J2000     & J2000       &/mJy&/$\mu$Jy                  &/$\mu$Jy        &&&$z$  &$z$\\
\hline
J033235$-$275530&03:32:35.09&$-$27:55:31.0&176.8\,$\pm$\,10.9&03:32:35.07&$-$27:55:32.6&4.7 &680\,$\pm$\,34$^{\star}$  &565\,$\pm$\,99  &$+$0.22&Y&0.038&0.051\\
J033229$-$274414&03:32:29.74&$-$27:44:14.4&156.7\,$\pm$\,10.9&03:32:29.87&$-$27:44:24.2&11.1&1,558\,$\pm$\,55$^{\star}$&1,670\,$\pm$\,84&$-$0.08&Y&0.077&0.077\\
J033250$-$273421&03:32:50.01&$-$27:34:21.3&159.3\,$\pm$\,11.1&03:32:50.41&$-$27:34:20.3&2.9 &470\,$\pm$\,20$^{\star}$  &706\,$\pm$\,56  &$-$0.49&N&0.251&0.251\\
J033128$-$273916&03:31:28.71&$-$27:39:16.1&105.3\,$\pm$\,11.1&03:31:28.82&$-$27:39:16.7&0.46&34.8\,$\pm$\,7.5          &$3\sigma<150$   &$-$0.80&N& --- &---\\
J033249$-$275842&03:32:49.47&$-$27:58:42.0&101.2\,$\pm$\,11.0&03:32:49.34&$-$27:58:44.7&0.32&216\,$\pm$\,16            &296\,$\pm$\,49  &$-$0.38&N& --- &2.215\\
J033124$-$275705&03:31:24.05&$-$27:57:05.5& 96.3\,$\pm$\,11.1&03:31:23.48&$-$27:56:58.5&1.3 &165\,$\pm$\,16            &183\,$\pm$\,52  &$-$0.12&N& --- &2.732\\
J033152$-$273931&03:31:52.82&$-$27:39:31.5& 93.4\,$\pm$\,11.0&03:31:52.07&$-$27:39:26.6&0.20&965\,$\pm$\,16            &765\,$\pm$\,52  &$+$0.28&Y& --- &2.296\\
J033258$-$274322&03:32:58.24&$-$27:43:22.3& 92.6\,$\pm$\,11.1&03:32:59.20&$-$27:43:25.1&0.53&148\,$\pm$\,17            &205\,$\pm$\,43  &$-$0.39&N& --- &1.160\\
J033129$-$275722&03:31:29.79&$-$27:57:22.6& 91.6\,$\pm$\,11.0&03:31:29.92&$-$27:57:22.4&0.27&144\,$\pm$\,16            &169\,$\pm$\,51  &$-$0.19&N& --- &1.571\\
J033319$-$275423&03:33:19.37&$-$27:54:23.2& 90.7\,$\pm$\,10.9&03:33:18.89&$-$27:54:33.8&0.37&60.2\,$\pm$\,15.0         &$3\sigma<150$   &$-$0.80&N& --- &0.488\\
J033246$-$275744&03:32:46.05&$-$27:57:44.0& 90.0\,$\pm$\,10.9&03:32:45.88&$-$27:57:44.8&2.7 &337\,$\pm$\,17$^{\star}$  &364\,$\pm$\,89  &$-$0.09&Y&0.103&0.105\\
J033149$-$274335&03:31:49.71&$-$27:43:35.9& 86.6\,$\pm$\,11.0&03:31:49.69&$-$27:43:26.4&0.95&192\,$\pm$\,16            &271\,$\pm$\,97  &$-$0.41&N&0.618&0.603\\
J033217$-$275905&03:32:17.07&$-$27:59:05.8& 85.3\,$\pm$\,11.0&03:32:16.99&$-$27:59:16.0&0.37&178\,$\pm$\,17            &395\,$\pm$\,45  &$-$0.96&N&0.126&0.220\\
J033318$-$274610&03:33:18.13&$-$27:46:10.3& 82.6\,$\pm$\,10.9&03:33:17.78&$-$27:46:05.9&0.43&100\,$\pm$\,14            &194\,$\pm$\,48  &$-$0.80&N&---  &2.059\\
J033216$-$280345&03:32:16.08&$-$28:03:45.0& 82.1\,$\pm$\,11.0&03:32:15.90&$-$28:03:47.1&0.77&192\,$\pm$\,20            &268\,$\pm$\,49  &$-$0.40&N&---  &0.452\\
J033145$-$274635&03:31:45.54&$-$27:46:35.5& 80.2\,$\pm$\,10.9&03:31:45.36&$-$27:46:40.2&0.29&42.2\,$\pm$\,7.3          &171\,$\pm$\,53  &$-$1.68&N&---  &---\\
J033308$-$274805&03:33:08.56&$-$27:48:05.9& 79.7\,$\pm$\,10.8&03:33:09.72&$-$27:48:01.3&0.20&433\,$\pm$\,19            &496\,$\pm$\,45  &$-$0.16&N&0.180&0.190\\
J033151$-$274431&03:31:51.15&$-$27:44:31.8& 74.0\,$\pm$\,10.9&03:31:51.09&$-$27:44:37.1&0.52&96.3\,$\pm$\,13           &$3\sigma<150$   &$-$0.53&Y&---  &1.911\\
J033130$-$275604&03:31:30.07&$-$27:56:04.5& 74.0\,$\pm$\,10.9&03:31:30.05&$-$27:56:02.2&0.90&1,270\,$\pm$\,64$^{\star}$&1,330\,$\pm$\,67&$-$0.06&Y&0.677&0.727\\
J033218$-$275216&03:32:18.03&$-$27:52:16.8& 73.6\,$\pm$\,10.9&03:32:18.25&$-$27:52:24.9&0.21&23.7\,$\pm$\,7.2          &$3\sigma<150$   &$-$0.80&N&0.739&0.724\\
J033240$-$280314&03:32:40.11&$-$28:03:14.7& 72.7\,$\pm$\,11.0&03:32:39.69&$-$28:03:10.5&0.34&168\,$\pm$\,30$^{\star}$  &707\,$\pm$\,180 &$-$1.73&N&---  &0.956\\
J033145$-$274205&03:31:45.02&$-$27:42:05.2& 72.8\,$\pm$\,11.0&03:31:44.46&$-$27:42:12.0&0.43&122\,$\pm$\,19            &181\,$\pm$\,45  &$-$0.47&N&---  &1.056\\
\hline
\end{tabular}
\end{center}
}

\vspace*{-2mm}
\noindent
Notes: $^{\dagger}$Flux densities before deboosting (see
\S\ref{seds}); these uncertainties include only instrumental noise
(\S\ref{blast}) and should be added in quadrature with the confusion
noise ($\sigma_{\rm confusion}=21.4$\,mJy -- \citealt{marsden09}) to
arrive at realistic SNRs; $^{\star}$Total flux density, determined
using {\sc tvstat}.  $^{\ddagger}$24-$\mu$m position coincident --
within 2\,arcsec -- of a {\em Chandra} source \citep{lehmer05,
luo08}. No further matches are made if the search radius is doubled.

\end{table*}

\section{Identifications}
\label{ids}

In order to avoid introducing a strong radio-related bias, and despite
the problems caused by the high surface density of 24-$\mu$m emitters
(evident in Fig.~\ref{postage}), we use the 24-$\mu$m counterpart
identifications of \citet{dye09} for our parent sample. Where
\citeauthor{dye09} does not present an association, or where they
favour a radio identification over a 24-$\mu$m source, we adopt our
own 24-$\mu$m identifications based on the probabilistic approach
outlined in the next section.

Before we discuss the 24-$\mu$m identifications -- and although we do
not use them hereafter -- we first describe the identification of
radio counterparts to the parent FIR catalogue. Radio imaging remains
the `gold standard' for pinpointing submm and FIR galaxies because of
the tight flux correlation and the low surface density of radio
emitters relative to optical/IR sources, and relative to the large
submm beam sizes. It is important, therefore, that we identify any
impending problems facing deep FIR surveys with {\em Herschel},
particularly in the SPIRE bands (250--500\,$\mu$m) where
\citet{devlin09} predict confusion limits only 1.5$\times$ lower than
those suffered by BLAST.

\subsection{Radio identifications}
\label{stats}

\begin{figure}
\begin{center}
  \includegraphics[scale=0.42,angle=270]{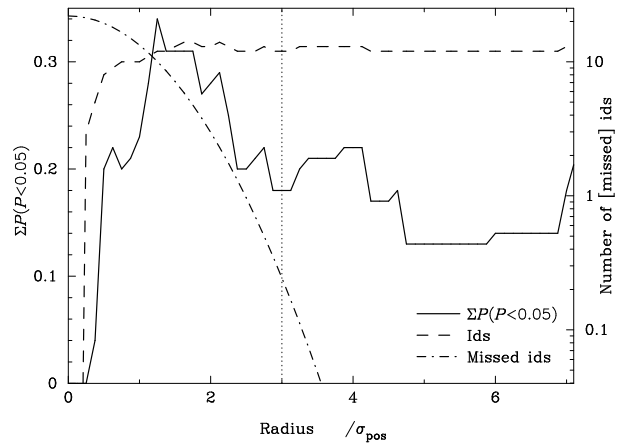}
  \caption{Illustration of how the integrated value of $P$ for
    identifications with $P<0.05$ (so $\Sigma P(P<0.05)$) for
    our sample varies with choice of search radius, where search
    radius is in multiples of $\sigma_{\rm pos}$ (the r.m.s.\ offset
    between observed and actual coordinates in R.A.\ or Dec.\ -- see
    equation~\ref{poseqn}).  This is the solid line, and the left-hand
    axis. In a situation not dominated by the effects of confusion,
    the value at the chosen radius equates directly to the expected
    number of spurious identifications amongst our sample. The
    right-hand axis shows the total number of identifications in our
    sample, and the number that we would expect to miss by selecting
    too small a search radius. The dotted line shows the adopted
    search radius -- in this case $3\,\sigma_{\rm pos}$ -- where we
    have 12 formal identifications and $\sim$0.24 was likely to have
    fallen outside our search radius.}
  \label{sump-vs-missed}
\end{center}
\end{figure}

The positional uncertainty of a source found in a map is given by

\begin{equation}
\sigma_{\rm pos} =
0.6\, [{\rm SNR}^2 - (2\gamma+4)]^{-1/2} \, {\rm FWHM}.
\label{poseqn}
\end{equation}

\noindent
for a situation where the counts obey a power law of the form
$N(>S)\propto S^{-\gamma}$ \citep{ivison07}. The SNR term here is
determined optimally in a beam-smoothed map and includes any variance
due to confusion. The signal is uncorrected for flux boosting and the
positional offset is not radial but is instead measured in Right
Ascension or Declination.

To associate radio emitters with the FIR sources in our sample, we
utilise the technique of \citet{downes86}, as implemented by
\citet{ivison02} and by most FIR/submm identification work thereafter.
According to \citeauthor{downes86}, $P$ is the probability of a random
association and depends on the surface density of radio emitters and
the radius within which one searches. The recent convention has been
to accept associations as secure identifications where $P<0.05$. We
assume a slope of 1.5 for the radio counts in the relevant flux
density regime ($\sim$40--400\,$\mu$Jy) and fine-tune the absolute
level of the counts with Monte Carlo simulations.

In choosing the search radius, we wish to maximise the number of
secure, unambiguous identifications and to minimise the number of real
counterparts missed. Fig.~\ref{sump-vs-missed} illustrates how the
total number of secure identifications and the integrated value of $P$
for those sources [hereafter $\Sigma P(P<0.05)$] varies with choice of
search radius. $\Sigma P(P<0.05)$ equates directly to the expected
number of spurious identifications in the sample. It is not
intuitively obvious why there is a well-defined peak in $\Sigma
P(P<0.05)$ when the total number of secure identifications is roughly
constant with search radius: we see this because faint, real radio
identifications compete (in terms of low $P$) with rare, bright
sources -- radio-loud AGN, entirely unrelated to the FIR source; as
the search radius increases, correct identifications drop out of the
integral (simply because their $P$ rises above 0.05); the resulting
reduction in secure identifications is balanced by radio-loud AGN --
contaminants; since they are very bright, their $P$ tends to zero as
soon as they are within the search radius; thus $\Sigma P(P<0.05)$
drops gently while the number of apparently `secure' identifications
is maintained. Allowing a search radius well beyond that of the peak
$P$ is therefore not advisable.

We express the search radius in terms of multiples of $\sigma_{\rm
  pos}$ to maintain an appropriate dependence on the SNR of the source
for which a counterpart is sought.  Based on
Fig.~\ref{sump-vs-missed}, we search within a radius of
3\,$\sigma_{\rm pos}$, though values as low as 1.5\,$\sigma_{\rm pos}$
are defensible choices. This yields 12 statistically secure
identifications with an expectation that $\sim$0.24 lie outside the
search area (calculated via the cumulative Rayleigh distribution
function).

At face value, the fraction of 250-$\mu$m sources with secure radio
identifications, 55 per cent, is similar to that found for SMGs
\citep[e.g.][]{smail00, ivison02, ivison07, pope06}, which are
expected to have higher luminosities and higher redshifts
\citep{chapman05}.  However, close examination of Fig.~\ref{postage},
in which we have superimposed the radio and 250-$\mu$m data on a
24-$\mu$m greyscale image of each BLAST source, reveals that our
ability to securely associate radio counterparts with FIR emitters has
been compromised. Broadly speaking, the BLAST sources are associated
with either large spirals or clumps of faint, radio and/or 24-$\mu$m
galaxies -- sometimes several of the former or as many as seven of the
latter (e.g.\ BLAST\,J033152$-$273931). At least half of the FIR
sources appear to be blends of two or more faint galaxies -- each
often with faint radio identifications -- i.e.\ confusion is a major
problem \citep{condon74, blain98}. This is compounded by the large
number of radio sources found by chance in the required search area (a
situation that worsens at 24\,$\mu$m, as mentioned earlier). Both
issues are related to the relatively large size of the 250-$\mu$m beam
(and the situation is worse at 350 and 500\,$\mu$m).

As an alternative approach, we inspected the data shown in
Fig.~\ref{postage} by eye, rejecting seven sources with complicated
250-$\mu$m morphologies suggestive of multiple, blended emitters. Of
the remaining fifteen, nine have statistically secure radio
identifications; of these, only three have unambiguous, single
identifications.  In total, therefore, only 14 per cent of our parent
sample have secure, unambiguous radio identifications, based on this
more subjective approach.

Thus, the purely statistical technique advocated previously for SMG
identification work proves unhelpful -- even misleading -- in the
confusion-dominated regime faced by BLAST and anticipated for the
deepest survey work with {\em Herschel} (see Fig.~\ref{blastmap}
caption). Working through the {\em Spitzer} and Photodetector Array
Camera and Spectrometer \citep[PACS --][]{poglitsch08} bands to
identify the dominant contributors to SPIRE flux densities may be more
fruitful -- but this has yet to be tested. This argues that a stacking
approach may be the most enlightening -- we turn to this later.

\begin{table*}
  \caption{FIR--submm photometry at the positions of the
    24-$\mu$m identifications, prior to deboosting, plus radio
    luminosities, temperatures and adopted redshifts.}
\label{multiband}
{\scriptsize
\begin{center}
\begin{tabular}{lcccccccccc}
\hline
Name&$S_{\rm 70\mu m}$&$S_{\rm 160\mu m}$&$S_{\rm 250\mu m}$&$S_{\rm 350\mu m}$&$S_{\rm 500\mu m}$&$S_{\rm 870\mu m}$&\td\ (obs)&\sfir                   &$L^{\alpha}_{\rm 1,400MHz}$&Adopted\\
    &/mJy              &/mJy              &/mJy              &/mJy            &/mJy            &/mJy             &/{\sc k}               &/10$^{-15}$\,W\,m$^{-2}$&/W\,Hz$^{-1}$&$z$\\
\hline
BLAST\,J033235$-$275530& 93.9\,$\pm$\,11.2&351.1\,$\pm$\,65.0&176.8\,$\pm$\,10.9&75.4\,$\pm$\,8.6&29.8\,$\pm$\,6.0  &1.23\,$\pm$\,1.16&25.2$^{+0.7}_{-0.9}$&8.28\,$\pm$\,1.79  &2.18\,$\times$\,10$^{21}$&{\em 0.038}\\
BLAST\,J033229$-$274414&136.6\,$\pm$\,15.4&368.0\,$\pm$\,65.8&141.6\,$\pm$\,10.8&71.7\,$\pm$\,8.5&36.8\,$\pm$\,6.0  &0.88\,$\pm$\,1.17&27.2$^{+0.9}_{-1.0}$&10.5\,$\pm$\,2.8   &2.13\,$\times$\,10$^{22}$&{\em 0.077}\\
BLAST\,J033250$-$273421& 30.1\,$\pm$\,4.9 &203.5\,$\pm$\,43.1&154.8\,$\pm$\,11.0&76.9\,$\pm$\,8.7&40.3\,$\pm$\,6.0  &1.82\,$\pm$\,1.22&21.5$^{+0.5}_{-0.9}$&4.22\,$\pm$\,1.24  &8.45\,$\times$\,10$^{22}$&{\em 0.251}\\
BLAST\,J033128$-$273916&  7.1\,$\pm$\,1.9 & 88.5\,$\pm$\,25.2&105.3\,$\pm$\,11.1&69.6\,$\pm$\,8.7&39.8\,$\pm$\,6.3  &4.02\,$\pm$\,1.18&18.1$^{+0.6}_{-0.8}$&---                &---                      &---  \\
BLAST\,J033249$-$275842&  8.4\,$\pm$\,2.1 & 72.6\,$\pm$\,21.7&101.2\,$\pm$\,10.9&66.4\,$\pm$\,8.6&22.6\,$\pm$\,6.0  &1.75\,$\pm$\,1.13&19.3$^{+0.7}_{-1.0}$&1.28\,$\pm$\,0.63  &6.38\,$\times$\,10$^{24}$&2.215\\
BLAST\,J033124$-$275705&  8.7\,$\pm$\,2.3 &100.7\,$\pm$\,30.1& 83.2\,$\pm$\,11.1&47.2\,$\pm$\,8.7&33.3\,$\pm$\,6.2  &2.99\,$\pm$\,1.20&19.0$^{+0.7}_{-0.9}$&1.40\,$\pm$\,0.70  &7.07\,$\times$\,10$^{24}$&2.732\\
BLAST\,J033152$-$273931&  3.0\,$\pm$\,1.4 & 31.9\,$\pm$\,14.4& 78.3\,$\pm$\,11.0&64.3\,$\pm$\,8.6&53.1\,$\pm$\,6.0  &3.49\,$\pm$\,1.14&15.9$^{+1.7}_{-1.8}$&0.77\,$\pm$\,0.42  &2.35\,$\times$\,10$^{25}$&2.296\\
BLAST\,J033258$-$274322&  7.4\,$\pm$\,2.0 & 82.8\,$\pm$\,24.8& 71.5\,$\pm$\,11.1&53.6\,$\pm$\,8.7&34.3\,$\pm$\,6.0  &4.31\,$\pm$\,1.17&18.3$^{+0.7}_{-0.9}$&1.21\,$\pm$\,0.61  &9.48\,$\times$\,10$^{23}$&1.160\\
BLAST\,J033129$-$275722&  3.6\,$\pm$\,1.7 & 49.0\,$\pm$\,19.9& 91.6\,$\pm$\,11.0&54.4\,$\pm$\,8.7&46.3\,$\pm$\,6.2  &5.11\,$\pm$\,1.14&16.1$^{+1.4}_{-1.0}$&0.93\,$\pm$\,0.49  &1.78\,$\times$\,10$^{24}$&1.571\\
BLAST\,J033319$-$275423&  3.9\,$\pm$\,1.4 & 36.0\,$\pm$\,14.9& 84.2\,$\pm$\,11.0&20.4\,$\pm$\,8.5&13.2\,$\pm$\,5.9  &1.12\,$\pm$\,1.13&18.8$^{+1.2}_{-1.4}$&0.71\,$\pm$\,0.41  &5.24\,$\times$\,10$^{22}$&0.488\\
BLAST\,J033246$-$275744& 46.0\,$\pm$\,6.0 &147.9\,$\pm$\,32.9& 90.0\,$\pm$\,10.9&46.7\,$\pm$\,8.5&18.7\,$\pm$\,5.9  &1.30\,$\pm$\,1.14&24.3$^{+1.0}_{-1.0}$&3.03\,$\pm$\,1.34  &8.49\,$\times$\,10$^{21}$&{\em 0.103}\\
BLAST\,J033149$-$274335& 15.7\,$\pm$\,2.8 &100.5\,$\pm$\,24.3& 81.5\,$\pm$\,11.0&46.6\,$\pm$\,8.6&30.9\,$\pm$\,5.9  &2.46\,$\pm$\,1.16&20.7$^{+0.8}_{-1.0}$&1.51\,$\pm$\,0.79  &2.75\,$\times$\,10$^{23}$&{\em 0.618}\\
BLAST\,J033217$-$275905& 39.5\,$\pm$\,5.9 &110.4\,$\pm$\,30.4& 80.5\,$\pm$\,11.0&63.6\,$\pm$\,8.5&30.4\,$\pm$\,6.0  &1.62\,$\pm$\,1.14&23.9$^{+1.1}_{-1.1}$&2.00\,$\pm$\,1.04  &7.17\,$\times$\,10$^{21}$&{\em 0.126}\\
BLAST\,J033318$-$274610&  5.6\,$\pm$\,1.7 & 62.3\,$\pm$\,20.8& 79.9\,$\pm$\,10.8&72.5\,$\pm$\,8.6&51.4\,$\pm$\,5.9  &3.84\,$\pm$\,1.13&17.7$^{+0.8}_{-0.9}$&1.00\,$\pm$\,0.56  &2.94\,$\times$\,10$^{24}$&2.059\\
BLAST\,J033216$-$280345& 16.0\,$\pm$\,3.2 & 85.7\,$\pm$\,24.5& 82.1\,$\pm$\,11.0&39.2\,$\pm$\,8.5&$-$0.3\,$\pm$\,6.0&2.04\,$\pm$\,1.30&21.2$^{+1.0}_{-1.1}$&1.31\,$\pm$\,0.70  &1.31\,$\times$\,10$^{23}$&0.452\\
BLAST\,J033145$-$274635&$-$0.3\,$\pm$\,1.0& 12.0\,$\pm$\,10.4& 75.2\,$\pm$\,10.8&11.0\,$\pm$\,8.5&12.8\,$\pm$\,5.9  &0.30\,$\pm$\,1.14&15.2$^{+1.1}_{-1.7}$&---                &---                      &---  \\
BLAST\,J033308$-$274805& 53.0\,$\pm$\,6.6 & 97.9\,$\pm$\,26.6& 66.1\,$\pm$\,10.8&37.0\,$\pm$\,8.5&31.7\,$\pm$\,5.9  &2.02\,$\pm$\,1.16&25.9$^{+1.3}_{-1.5}$&1.99\,$\pm$\,1.04  &3.65\,$\times$\,10$^{22}$&{\em 0.180}\\
BLAST\,J033151$-$274431&  6.0\,$\pm$\,1.7 & 77.5\,$\pm$\,20.2& 74.0\,$\pm$\,10.8&63.8\,$\pm$\,8.5&37.4\,$\pm$\,5.9  &4.18\,$\pm$\,1.16&17.9$^{+0.7}_{-0.9}$&1.05\,$\pm$\,0.61  &2.13\,$\times$\,10$^{24}$&1.911\\
BLAST\,J033130$-$275604& 28.5\,$\pm$\,4.1 & 83.2\,$\pm$\,24.7& 74.0\,$\pm$\,10.8&20.9\,$\pm$\,8.5&44.9\,$\pm$\,6.0  &3.05\,$\pm$\,1.15&23.0$^{+1.1}_{-1.2}$&1.61\,$\pm$\,0.91  &2.10\,$\times$\,10$^{24}$&{\em 0.677}\\
BLAST\,J033218$-$275216&  5.3\,$\pm$\,1.4 & 54.2\,$\pm$\,17.8& 71.4\,$\pm$\,10.8&52.6\,$\pm$\,8.5&46.8\,$\pm$\,5.9  &7.74\,$\pm$\,1.17&16.1$^{+2.3}_{-1.1}$&0.88\,$\pm$\,0.52  &5.63\,$\times$\,10$^{22}$&{\em 0.739}\\
BLAST\,J033240$-$280314&  6.8\,$\pm$\,2.0 & 50.7\,$\pm$\,18.6& 72.7\,$\pm$\,11.0&61.8\,$\pm$\,8.5&30.6\,$\pm$\,6.1  &4.05\,$\pm$\,1.25&18.2$^{+1.4}_{-1.4}$&0.90\,$\pm$\,0.54  &9.53\,$\times$\,10$^{23}$&0.956\\
BLAST\,J033145$-$274205&  4.8\,$\pm$\,1.6 & 45.0\,$\pm$\,17.2& 56.0\,$\pm$\,11.0&27.0\,$\pm$\,8.7&28.4\,$\pm$\,6.2  &1.98\,$\pm$\,1.14&18.8$^{+1.0}_{-1.5}$&0.70\,$\pm$\,0.45  &6.38\,$\times$\,10$^{23}$&1.056\\
 \hline
\end{tabular}
\end{center}
}

\vspace*{-2mm}
\noindent
Notes: $L^{\rm \alpha}_{\rm 1,400MHz}$ is the $K$-corrected radio
luminosity using the measured $\alpha$. \sfir\ is deboosted; this
process dominates the error budget (see \S\ref{blast} and
\S\ref{seds}). Spectroscopic redshifts are shown in italics.

\end{table*}

\subsection{24-$\mu$m identifications}
\label{24um}

Using a different probabilistic approach to our own, \citet{dye09}
listed the most likely 24-$\mu$m counterparts for many of our
250-$\mu$m-selected galaxies. To their list, we added
BLAST\,J033229$-$274414, an obvious blend of two galaxies at $z\sim
0.08$ (Fig.~\ref{postage}), and in a further six cases where
\citeauthor{dye09} listed no association we added the most likely of
our own 24-$\mu$m identifications. We list the 24-$\mu$m positions in
Table~\ref{squad}, together with the radio properties at those
positions. The FIR and submm properties at the 24-$\mu$m positions are
given in Table~\ref{multiband}.

As already noted, we adopt these 24-$\mu$m counterparts to avoid
introducing a strong radio-related bias, and despite the problems
caused by the high surface density of 24-$\mu$m emitters, exacerbated
by FIR confusion.

\begin{figure}
\begin{center}
  \includegraphics[scale=0.43,angle=270]{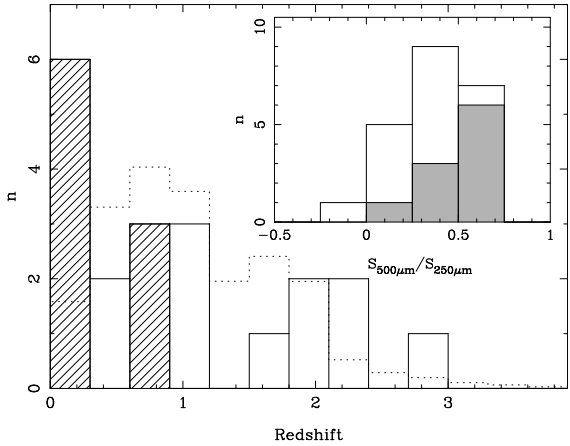}
  \caption{The redshift distribution of the most likely 24-$\mu$m
    associations with our BLAST 250-$\mu$m sample. The dotted line
    shows the redshift distribution of the 24-$\mu$m sources used for
    stacking in \S\ref{stacking}. The subset of galaxies with
    spectroscopic redshifts are shown as a hatched area. Inset is a
    histogram of $S_{\rm 500\mu m}/S_{\rm 250\mu m}$, where we expect
    galaxies at higher redshifts to have higher ratios. Sources at
    redshifts higher than the median ($z=0.74$) are shaded and there
    is a very clear tendency for these to have higher ratios, as we
    would expect.}
  \label{nz}
\end{center}
\end{figure}

\begin{figure}
\begin{center}
  \includegraphics[scale=0.43,angle=270]{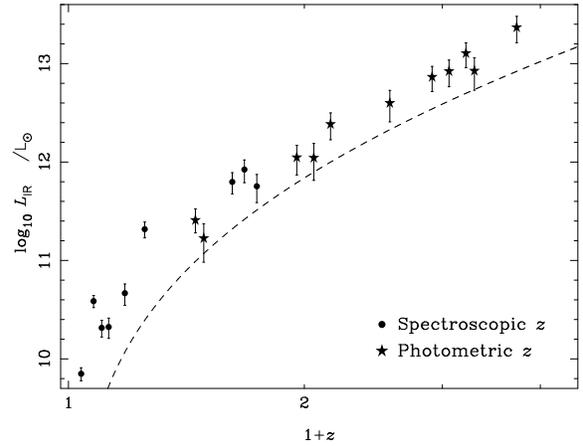}
  \caption{Deboosted IR luminosity as a function of redshift. The
    dashed line is an arbitrarily scaled detection threshold for our
    adopted cosmology. Galaxies with spectroscopic and photometric
    redshifts are plotted as circles and stars, respectively.}
\label{firz}
\end{center}
\end{figure}

\begin{figure}
\begin{center}
  \includegraphics[scale=0.43,angle=270]{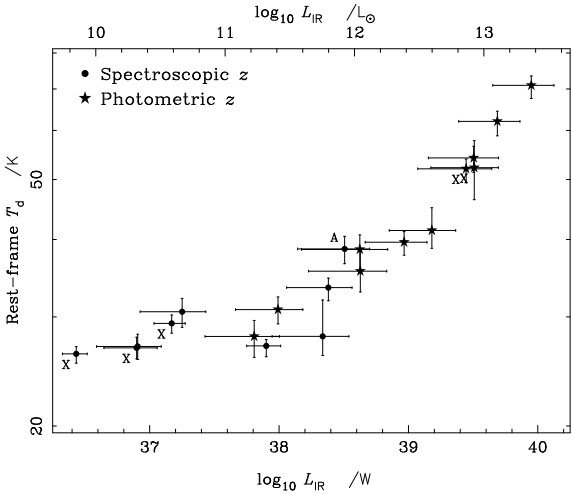}
  \caption{\td\ versus deboosted IR luminosity. Galaxies with
    spectroscopic and photometric redshifts are plotted as circles and
    stars, respectively. A radio-loud AGN, selected via its radio
    morphology, is labelled `A'; X-ray emitters are labelled `X'. The
    strong correlation between \td\ and $L_{\rm 60\mu m}$ seen by
    \citet{dunne00} is also evident here. Selection at 250\,$\mu$m
    tends to find local galaxies with \td\ $\approx$ 20\,{\sc k}, but
    inevitably also catches more luminous, distant galaxies with the
    same {\em observed} \td\ and {\em rest-frame} values of \td\ that
    are $\approx (1 + z)\times$ higher. }
\label{td-lfir}
\end{center}
\end{figure}

\section{Results}
\label{results}

\subsection{Redshift distribution}
\label{nzwords}

Of the most likely 24-$\mu$m counterparts for our 250-$\mu$m sample,
eight have spectroscopic redshifts \citep[see][]{croom01, lefevre04,
szokoly04, vanzella08, eales09} and another 11 have photometric
redshifts \citep[][in preparation]{rafferty09}.  For every
spectroscopic case, both types of estimate are available, and these
agree reasonably well ($\delta z_{\rm r.m.s.}\ls 0.05$, with one
outlier).  We were unable to assign redshifts with confidence in two
cases (see Tables~\ref{squad} and \ref{multiband}).

The redshift distribution of 24-$\mu$m-identified, 250-$\mu$m-selected
sources is shown in Fig.~\ref{nz}. The median is $z$ = 0.74 with an
interquartile range of $z$ = 0.25--1.57, so slightly beyond the
70-$\mu$m-selected sample of \citet{symeonidis09} which had a median
$z$ of 0.42. Inset in the same plot is a histogram of $S_{\rm 500\mu
m}/S_{\rm 250\mu m}$. This ratio should increase with redshift as the
SED peak shifts to longer wavelengths and we find a strong tendency
for those above the median redshift to have higher ratios.

\subsection{Far-infrared SEDs}
\label{seds}

Flux boosting of nearly the same magnitude as that experienced in the
BLAST wavebands (\S\ref{blast}) also influences our {\em Spitzer}
measurements and, to a lesser extent, those made using LABOCA. We
therefore apply deboosting factors appropriate for the BLAST
250-$\mu$m sources -- appropriate to the level of precision required
\citep{eales09} -- to \sfir, which we define to be the flux observed
between rest-frame 8 and 1,000\,$\mu$m. The large uncertainty in the
deboosting factor is propagated in quadrature with other
uncertainties.  \sfir\ could be determined via modified blackbody fits
to the 70-, 160-, 250-, 350-, 500- and 870-$\mu$m flux measurements
with $\beta$ fixed to 1.5 (for example). However, with such
well-sampled SEDs at our disposal, we concluded that \sfir\ is better
determined by interpolating between the measured flux densities at
24--870\,$\mu$m, and extrapolating outside the data range to
rest-frame 8 and 1,000\,$\mu$m with spectral indices of $-1.8$
(appropriate for an M\,82-like SED) and $+3.5$, respectively, then
deboosting according to the recipe in \citet{eales09}.  These are the
values listed in Table \ref{multiband}. Fig.~\ref{firz} shows the
deboosted IR luminosity as a function of redshift for our sample.

We do use modified blackbody fits to estimate the effective dust
temperature, \td, and these are also presented in
Table~\ref{multiband}. We note that several dust components with a
range of temperatures would be required to replicate the SEDs
accurately over the full wavelength range. The average observed-frame
\td\ is 20.5\,$\pm$\,3.5\,{\sc k}. Fig.~\ref{td-lfir} shows the
relationship between rest-frame \td\ and \lfir. We expect and see a
strong correlation, following the work by \citet{dunne00} and
\citet{chapman05}: at 250\,$\mu$m we are sensitive to a particular
observed temperature regime; rest-frame \td\ is higher (by a factor 1
+ $z$) than observed \td, so for more distant detected sources we
expect hotter dust. \lfir\ scales with a high power of \td, yielding
the strong correlation evident in Fig.~\ref{td-lfir}.

\subsection{The FIR/radio correlation}
\label{fir-radio}

We have explored trends in the FIR/radio correlation, both the
monochromatic version, e.g., $q_{70} = {\rm log}_{10} \, (S_{\rm 70\mu
  m}/S_{\rm 1,400MHz})$ as defined by \citet{appleton04} -- where
$K$-correcting both $S_{\rm 70\mu m}$ and $S_{\rm 1,400MHz}$ requires
assumptions about SED shape if a significant range of redshift is
present -- and the correlation between bolometric \sfir\ and
$K$-corrected radio flux density, where both quantities are well
constrained.

We adopt a standard routine so that we can compare like with like.
First, we exclude a radio-loud AGN, identified on the basis of its
radio morphology (BLAST\,J033130$-$275604). We do not exclude the
X-ray emitters (see Table~\ref{squad}). Next, we calculate the
error-weighted mean and standard deviation of the relevant $q$
parameter, noting points that deviate from the mean by more than
3$\sigma$ (these being `radio-excess AGN' candidates, as described by
\citealt{donley05}). The resultant statistics (and notes) are reported
in Table~\ref{qtab}.

\subsubsection{Monochromatic correlations}

\begin{figure}
\begin{center}
  \includegraphics[scale=0.43,angle=270]{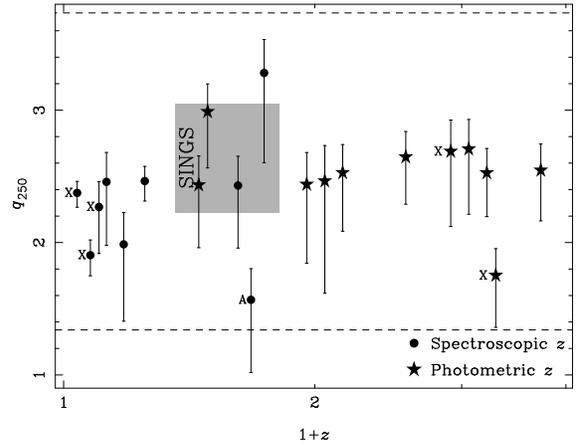}
  \caption{\qtwofifty\ versus redshift, where no $K$ corrections have
    been applied. Circles and stars represent spectroscopic and
    photometric redshifts, respectively. A radio-loud AGN, selected
    via its radio morphology, is labelled `A'; X-ray emitters are
    labelled `X'.  The shaded region (around $z\sim 0.56\pm 0.20$,
    where $\lambda_{\rm rest}$ = 160\,$\mu$m for the BLAST 250-$\mu$m
    filter) represents the $\pm\sigma$ range of $q_{\rm 160}$ measured
    for SINGS \citep{dale07}. A minor correction has been made to
    $q_{\rm 160}$ ($-$0.04) to emulate the effect of $K$-correcting
    the $z=0.56$ radio flux density for the BLAST sample. $\pm3\sigma$
    deviations from the mean value of \qtwofifty\ are shown by dashed
    lines. }
\label{q250}
\end{center}
\end{figure}

The monochromatic relationship between $S_{\rm 250 \mu m}$ and $S_{\rm
  1,400MHz}$, \qtwofifty\ = ${\rm log}_{10} \, (S_{\rm 250\mu
  m}/S_{\rm 1,400MHz})$ is shown in Fig.~\ref{q250} as a function of
  redshift, and the resultant statistics are reported in
  Table~\ref{qtab}. Also shown as a grey box in Fig.~\ref{q250} is the
  $\pm 1 \sigma$ range of $q_{\rm 160}$ measured for the 57 members of
  the {\em Spitzer} IR Nearby Galaxies Survey (SINGS) sample which are
  detected at both 160\,$\mu$m and 1,400\,MHz \citep{dale07}. At
  $z\approx\rm 0.6$, where the BLAST 250-$\mu$m filter samples
  rest-frame 160-$\mu$m emission, values of $q_{\rm 160}$ and $q_{\rm
  250}$ are entirely compatible with the SINGS average, with or
  without the small radio $K$ correction.  We interpret this as
  evidence for a low rate of evolution, at least out to $z\sim 0.6$.

\begin{figure}
\begin{center}
  \includegraphics[scale=0.43,angle=270]{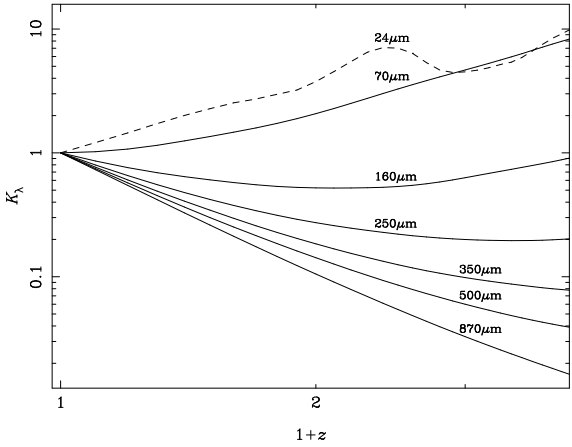}
  \caption{$K$ corrections for the {\em Spitzer}, BLAST and LABOCA
    wavebands, assuming an M\,82 SED template, calculated using accurate
    filter transmission profiles and defined such that $K =
    S^K_{\nu}/S_{\nu}^{\rm obs}$, so excluding the $(1+z)$ term that
    takes care of the extra width of the differential frequency
    element in the observed frame (equation~\ref{keqn}).}
\label{kcorr}
\end{center}
\end{figure}

\begin{table}
\caption{Summary of $q$ statistics.}
\label{qtab}
{\scriptsize
\begin{center}
\begin{tabular}{clcl}
\hline
Type  &\multicolumn{1}{c}{Mean}&Standard &$K$ correction(s)\\
of $q$&       &deviation&or stacking parameters\\
\hline
\multicolumn{4}{l}{250-$\mu$m-selected galaxy sample:}\\
\qtwofifty & 2.26 & 0.35  &None\\
\qtwofifty & 1.85$^{\star}$ & 0.43  &M\,82 SED, $\alpha=-0.8$\\
\qtwofifty & 1.82$^{\star}$ & 0.43  &Arp\,220 SED, $\alpha=-0.8$\\
\qtwofifty & 2.08 & 0.34  &Fig.~\ref{best} SED, $\alpha=-0.8$\\
\qtwofifty & 1.94 & 0.34  &M\,82 SED, measured $\alpha$\\
\qtwofifty & 1.91 & 0.34  &Arp\,220 SED, measured $\alpha$\\
\smallskip
\qtwofifty & {\bf 2.18} & {\bf 0.28}  &Fig.~\ref{best} SED, measured $\alpha$\\
\qfir      & 2.40 & 0.29  &None\\
\qfir      & 2.37 & 0.31  &$\alpha=-0.8$\\
{\bf \qfir}& {\bf 2.41} & {\bf 0.20}  &Measured $\alpha$\\
\hline
\multicolumn{4}{l}{SINGS sample \citep{dale07}:}\\
$q_{\rm 160}$&2.68 & 0.41 &None\\
\hline
\multicolumn{4}{l}{Stacking on 24-$\mu$m-selected galaxies:}\\
\qtwofifty & 2.70 & 0.08  &None\\
\qtwofifty & 2.00 & 0.37  &M\,82 SED, measured $\alpha$\\
\smallskip
\qtwofifty & 2.31 & 0.18  &Fig.~\ref{best} SED, measured $\alpha$\\
\qfir      & 2.89 & 0.06  &None\\
\qfir      & 2.67 & 0.17  &Measured $\alpha$\\
\hline
\multicolumn{4}{l}{Stacking on radio-selected galaxies (no $K$ correction):}\\
\qtwofifty & 2.05  & 0.34   &Measured $\alpha$\\
\qtwofifty & 2.42  & --     &$40<S_{\rm 1,400MHz}<100\,\mu$Jy [507]\\
\qtwofifty & 2.17  & --     &$100<S_{\rm 1,400MHz}<251\,\mu$Jy [224]\\
\qtwofifty & 1.81  & --     &$251<S_{\rm 1,400MHz}<631\,\mu$Jy [69]\\
\smallskip
\qtwofifty & 1.56  & --     &$631<S_{\rm 1,400MHz}<1,000\,\mu$Jy [16]\\
\qfir      & 1.96  & 0.34   &Measured $\alpha$\\
\qfir      & 2.25  & --     &$40<S_{\rm 1,400MHz}<100\,\mu$Jy [507]\\
\qfir      & 2.11  & --     &$100<S_{\rm 1,400MHz}<251\,\mu$Jy [224]\\
\qfir      & 1.87  & --     &$251<S_{\rm 1,400MHz}<631\,\mu$Jy [69]\\
\qfir      & 1.29  & --     &$631<S_{\rm 1,400MHz}<1,000\,\mu$Jy [16]\\
\hline
\end{tabular}
\end{center}
}

\noindent
$^{\dagger}$ With M\,82-like power-law SED ($\alpha = -1.8$)
             shortward of the peak;

\noindent
$^{\star}$  One radio-excess AGN present.
\end{table}

\subsubsection{$K$-corrected monochromatic correlations}

Next, we looked at the effect of `$K$ correcting' the monochromatic
relationship between 250-$\mu$m and 1,400-MHz flux densities, where $K$
is the function that transforms a quantity observed at some frequency into an
equivalent measurement in the rest frame of the source being
observed. Taking the simple case of a power-law SED, where $L^K_{\nu}$
and $S^{\rm obs}_{\nu}$ are the $K$-corrected luminosity density and
observed flux density, respectively, and $d_{\rm lum}$ is the
luminosity distance (in m),

\begin{equation}
L^K_{\nu} = \frac{4 \pi d_{\rm lum}^2}{(1+z)} \, K \, S^{\rm obs}_{\nu} ,
\label{keqn}
\end{equation}

\noindent
where the $(1+z)$ term takes care of the extra width of the
differential frequency element in the observed frame. Now imagine we
are working on a sample of $z=1$ galaxies at 1,400\,MHz: the factor $K$
converts $S^{\rm obs}_{\rm 1,400MHz}$ -- radiation emitted at 2,800\,MHz
and seen at 1,400\,MHz -- into $S^K_{\rm 1,400MHz}$, the 1,400-MHz flux
density in the rest frame of the emitter.  Using the definitions of
$z$ and spectral index, $\alpha$, i.e.\ $(1+z)=\nu_{\rm rest}/\nu_{\rm
obs}$ and $S_{\nu}\propto \nu^{\alpha}$, we find that $K =
S^K_{\nu}/S^{\rm obs}_{\nu} = (1+z)^{-\alpha}$, so

\begin{equation}
L^K_{\nu} = \frac{4 \pi d_{\rm lum}^2}{(1+z)} \, S^{\rm obs}_{\nu}
\, (1+z)^{-\alpha} = 4 \pi d_{\rm lum}^2 \, \frac{S^{\rm obs}_{\nu}}
{(1+z)^{1+\alpha}} .
\end{equation}

\begin{figure}
\begin{center}
  \includegraphics[scale=0.4,angle=0]{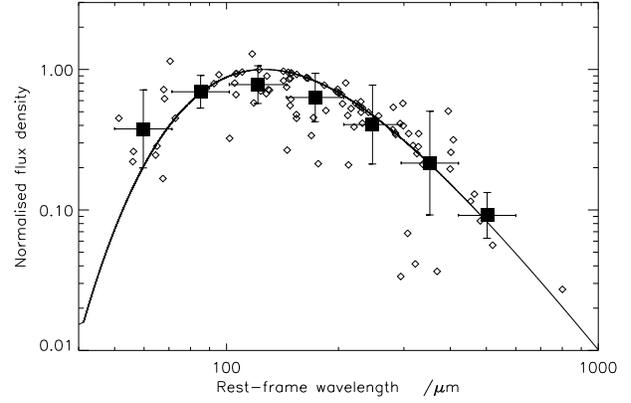}
  \caption{Template SED with arbitrary flux scale built from the
    normalised photometry of those 250-$\mu$m-selected galaxies with
    secure identifications using the technique described by
    \citet{pope06}. The solid line is a 36-{\sc k} modified
    blackbody. To create a template from this SED, suitable for use in
    a $K$-correction algorithm, we interpolated between the means then
    extrapolated to shorter and longer wavelengths with $\alpha=-1.8$
    and $+2.25$, respectively.}
  \label{best}
\end{center}
\end{figure}

For the radio spectrum we use the measured slope between 610 and
1,400\,MHz, adopting $\alpha = -0.8$ where this is consistent with the
610-MHz upper limit, or where no measurement is available.  This
results in a median $\alpha$ of $-$0.4. In the FIR waveband, we
explored two options.  One is to adopt the observed SED of M\,82,
which \citet{ibar08} found gave the least scatter in $q_{24}$. $K$
corrections\footnote{Our {\sc idl}-based $K$-correction code, together
  with SED templates and filter profiles, are available on request.}
appropriate for an M\,82 SED template in the {\em Spitzer}, BLAST and
LABOCA wavebands, calculated using accurate filter transmission
profiles, are shown in Fig.~\ref{kcorr}. We also tried the SED of
Arp\,220 and an SED built from those 250-$\mu$m-selected galaxies with
secure identifications and redshifts, normalised at 250\,$\mu$m using
the technique of \citet{pope06} -- see Fig.~\ref{best}. The latter SED
made an appreciable difference to the absolute value of \qtwofifty\
and resulted in less scatter. Relevant statistics for $q_{\rm 250}$
under each set of assumptions are reported in Table~\ref{qtab}.

\begin{figure}
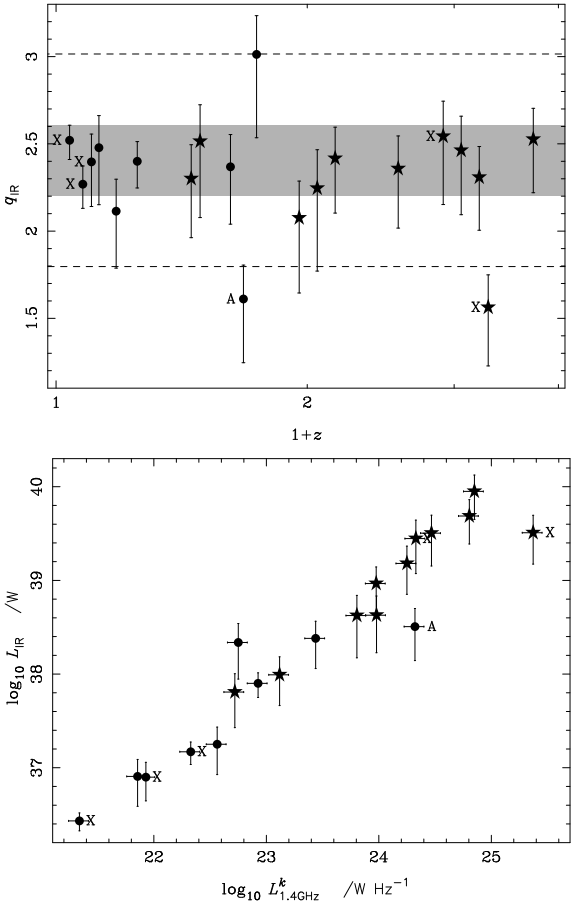

\begin{center}
  \includegraphics[scale=0.43,angle=270]{qbol-upper.eps}
  \vspace*{0.2cm}

  \includegraphics[scale=0.43,angle=270]{qbol-lower.eps}
  \caption{{\em Upper:} \qfir\ as a function of redshift, using
    $K$-corrected radio luminosities based on measured radio
    spectra. The shaded area represents $\pm\sigma$; dashed lines are
    at $\pm3\sigma$. {\em Lower:} $L_{\rm IR}$ versus $L^{\alpha}_{\rm
    1,400MHz}$. A radio-loud AGN, selected via its radio morphology,
    is labelled `A'; X-ray emitters are labelled `X'. }
  \label{qbolk}
\end{center}
\end{figure}

\subsubsection{$K$-corrected bolometric correlations}

Moving from the monochromatic correlation to bolometric quantities,
where a $K$ correction is required only in the radio waveband.
 \citet{helou85} defined \qfir\ such that

\begin{equation}
q_{\rm IR} =
{\rm log} \left( \frac{S_{\rm FIR}/3.75\times10^{12}}{\rm W\,m^{-2}} \right) -
{\rm log} \left( \frac{S_{\rm 1,400MHz}}{\rm W\,m^{-2}\,Hz^{-1}} \right) ,
\end{equation}

\noindent
where they defined $S_{\rm FIR}$ as the flux between 42.5 and
122.5\,$\mu$m and $3.75\times10^{12}$ is the frequency at their
mid-band, 80\,$\mu$m. Working at 42.5--122.5\,$\mu$m would exploit
only one tenth of the wavelength coverage now at our disposal so,
although we adopt this definition, we use \sfir\ -- the flux between
rest-frame 8 and 1,000\,$\mu$m -- instead of $S_{\rm FIR}$.

We $K$-correct the radio flux densities as described in the discussion
of monochromatic relations.  The results are shown in Fig.~\ref{qbolk}
and reported in Table~\ref{qtab}. Utilising all the data at our
disposal, we find \qfir\ = $2.41\pm 0.20$ for our 250-$\mu$m-selected
galaxy sample, with no evidence of redshift dependence.

\subsection{Assessing redshift evolution of the FIR/radio correlation by stacking}
\label{stacking}

\citet{marsden09}, \citet{pascale09} and \citet{patanchon09} have
demonstrated the power of stacking analyses based on confusion-limited
maps and the very steep BLAST source counts.  We have applied those
same techniques to study the evolution of \qtwofifty\ and \qfir\ by
stacking at the positions of the thousands of mid-IR-selected galaxies
with photometric redshifts in ECDFS. We thus eliminate the biases
introduced by starting from a flux-boosted and confused FIR-selected
sample and the problems of demanding secure identifications in a
densely populated 24-$\mu$m image.  Since \citet{marsden09} has shown
that the BLAST catalogue only comprises $\approx$10 per cent of the
total flux in the maps, stacking, which uses all of the pixels, has
the potential to greatly increase the sensitivity of the analysis.

However, some new problems are introduced. We have to worry about
selection effects due to the catalogue of positions upon which we
stack -- in this case a 24-$\mu$m-selected sample, with flux densities
determined using IRAC positions as priors \citep{magnelli09}, with the
redshift distribution shown in Fig.~\ref{nz} and with a non-trivial
$K$ correction (see Fig.~\ref{kcorr}).  \S5 of \citet{pascale09}
provides a practical demonstration of the paradigm described by
\citet{marsden09} and shows that our fundamental approach to stacking
is accurate to the level of precision required here.

\begin{figure}
\begin{center}
  \includegraphics[scale=0.43,angle=270]{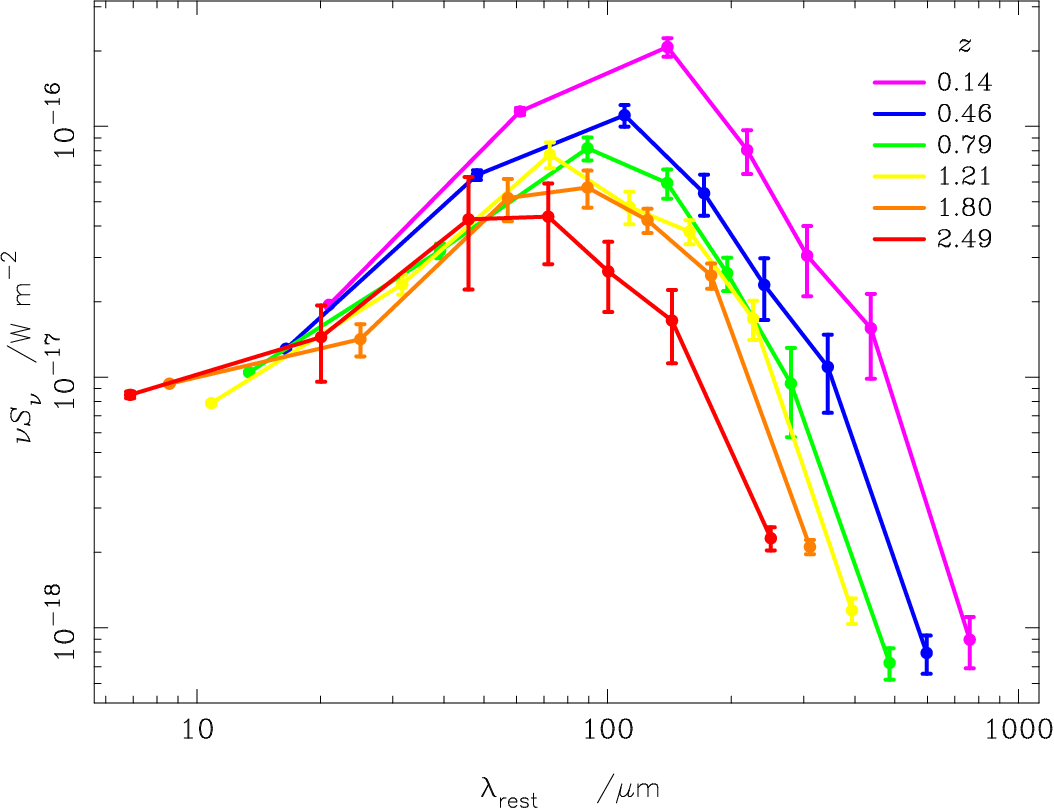}
  \vspace*{0.2cm}

  \includegraphics[scale=0.43,angle=270]{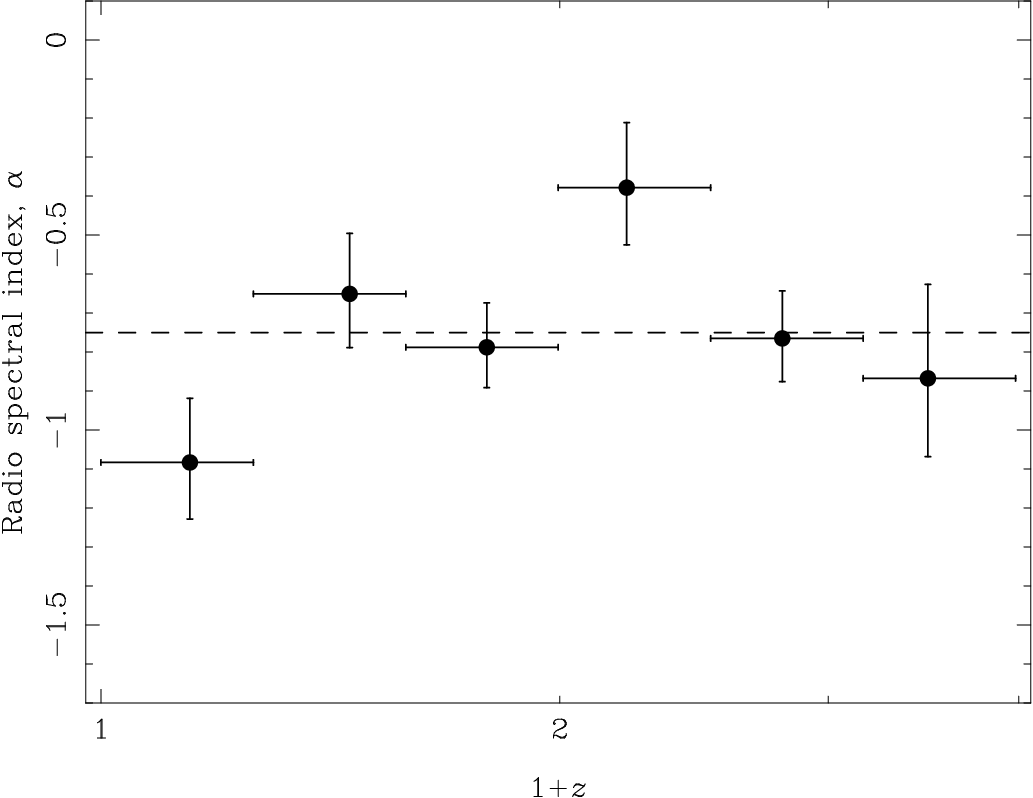}
  \vspace*{0.2cm}

  \includegraphics[scale=0.43,angle=270]{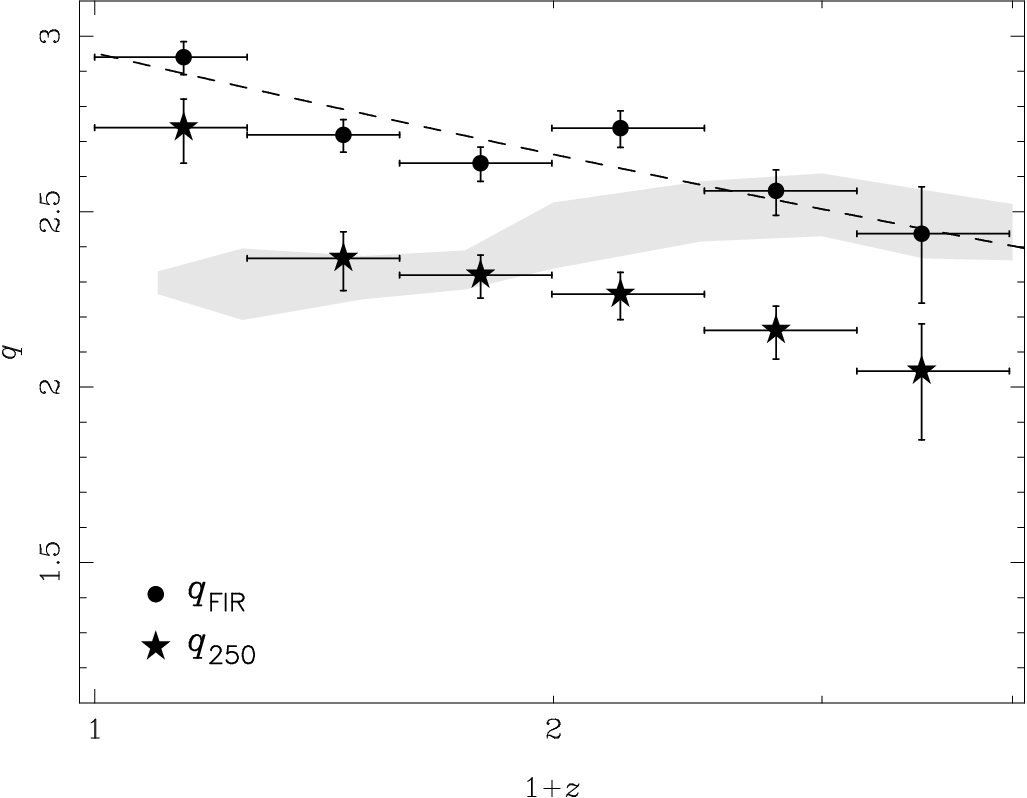}
  \caption{{\em Top:} $\nu_{\rm obs} S_{\nu_{\rm obs}}$ versus rest
    wavelength in six redshift bins, measured at the positions of the
    24-$\mu$m sample \citep{magnelli09, pascale09}.  {\em Middle:}
    Radio spectral index, $\alpha$, in the six redshift bins.  The
    mean value is $-0.75\pm 0.06$ and there is no evidence of
    significant evolution across $0<z<3$. {\em Bottom:} \qtwofifty\
    and \qfir\ in the six redshift bins. Dashed line: weighted
    least-squares fit of the form \qfir\ $\propto (1+z)^{-0.15}$;
    shaded area: $\pm 1\sigma$ prediction of the \citet{lacey08}
    galaxy-formation model and the spectrophotometric model of
    \citet{silva98} for galaxies with $L_{\rm
      IR}>10^{12}$\,L$_{\odot}$, as implemented by
    \citet{swinbank08}, updated to reflect our wavelength coverage and
    definition of \qfir. $K$-corrections are based on the measured
    radio spectra and -- for \qtwofifty\ only -- the SED discussed in
    \S\ref{fir-radio} and shown in Fig.~\ref{best}. The scale is
    chosen to match Fig.~\ref{radio-stack}.}
  \label{stack}
\end{center}
\end{figure}

Median-stacking in the radio regime should reduce the consequences of
radio-loud AGN becoming more prevalent at high redshift
\citep{dunlop90}.  Stacking should, therefore, allow us to probe the
evolution of $q$ for galaxies more representative of the general
population than is possible for a sample requiring both FIR and radio
detections, modulo 24-$\mu$m selection biases.

We adopt the technique of \citet{ivison07b}, dividing the
\citeauthor{magnelli09} catalogue into six redshift bins, evenly
spaced in log$_{10} \, (1 + z)$ between $z$ = 0 and 3, containing 561,
1,441, 2,205, 1,823, 1,900 and 372 galaxies, respectively.
Error-weighted mean postage-stamp images are obtained in all the FIR
filters, allowing us to assess the level of any background
emission. Median images are made from the VLA and GMRT stacks, to
reduce the influence of radio-loud AGN. In the radio regime, where the
spatial resolution is relatively high, making images allows us to
conserve flux density that would otherwise be lost due to smearing by
astrometric uncertainties and finite bandwidth (chromatic aberration)
at the cost of larger flux density uncertainties. Total flux densities
are measured using Gaussian fits within \AIPS; we determine the
appropriate source centroid and width by making radio stacks using
every available 24-$\mu$m source (SNR $\sim$ 50), then fix these
parameters to minimise flux density uncertainties and avoid spurious
fits; at 1,400\,MHz, where both forms of smearing are most prevalent,
flux densities were $\sim$2$\times$ higher than the peak values. Such
losses are not expected where the astrometry is accurate to a small
fraction of a beam, as it is at 70--870\,$\mu$m.

The {\em top} and {\em middle} panels of Fig.~\ref{stack} show the
evolution with redshift of the overall SED and the median radio
spectral index, $\alpha$, for the \citet{magnelli09} sample. We see
that the rest-frame 8--100-$\mu$m portion of the SED becomes
progressively more important -- in terms of its relative contribution
to \sfir\ -- as we move to higher redshifts, despite the 870-$\mu$m
flux density increasing slightly with redshift \citep[as expected
--][]{blain93}. We find no evidence of significant deviations from the
error-weighted mean spectral index, $\alpha^{\rm 1,400}_{\rm
  610}=-0.75\pm 0.06$, nor of a significant trend in $\alpha^{\rm
  1,400}_{\rm 610}$ across $0<z<3$. A weighted least-squares fit
yields $\alpha^{\rm 1,400}_{\rm 610}\propto (1+z)^{0.14\pm0.20}$.

The lower panel of Fig.~\ref{stack} shows the evolution with redshift
of \qtwofifty\ and \qfir, the former $K$-corrected using the SED
template described in \S\ref{fir-radio} and shown in Fig.~\ref{best},
both $K$-corrected using measured radio spectra. Mean values of both
\qtwofifty\ and \qfir\ lie within 1$\sigma$ of the mean values found
for our 250-$\mu$m-selected galaxies (Table~\ref{qtab}).  We find,
however, that \qfir\ evolves with redshift, even across $0<z<1$ where
incompleteness should not be a major issue in our 24-$\mu$m sample. A
weighted least-squares fit suggests \qfir\ $\propto
(1+z)^{-0.15\pm0.03}$.  \qtwofifty\ is also found to evolve, though
the form of the evolution is strongly dependent on the choice of SED
adopted for $K$ correction, i.e.\ on the shape of the template SED
shortward of 200\,$\mu$m, emphasising the importance of good spectral
coverage. The evolution is even stronger for M\,82 and Arp\,220
SEDs. The different behaviour with respect to redshift seen in
Figs~\ref{qbolk} and \ref{stack} is presumably due to selection
biases: the lack of evolution seen in Fig.~\ref{qbolk} reflects the
FIR sample selection employed there; the evolution seen in
Fig.~\ref{stack} reflects the IR and radio characteristics of the
increasingly luminous 24-$\mu$m emitters upon which we are stacking at
progressively higher redshifts.

The use of {\it median} radio images should exclude radio-loud AGN at
least as effectively as the 3-$\sigma$ clip employed in
\S\ref{fir-radio}, so the influence of the anticipated evolution (with
redshift) of radio-loud AGN on the stacked values of $q$ should be
minimised. Using means instead, which we view as an inferior approach,
the average spectral index, $\alpha^{\rm 1,400}_{\rm 610}$, falls to
$-1.12$ as would be expected if steep-spectrum AGN contaminate the
stacks. This, together with the $\sim$3$\times$ higher mean radio flux
densities, shifts \qfir\ by $-0.67$, although the form of the redshift
evolution remains similar, with \qfir\ $\propto(1+z)^{-0.14\pm 0.04}$.

Responding to reports that the radio background is significantly
brighter than the cumulative intensity seen in discrete radio emitters
\citep{fixsen09, seiffert09}, \citet{singal09} have speculated that we
should see the FIR/radio correlation evolve, driven by
$\sim$0.01--10-$\mu$Jy radio activity amongst ordinary star-forming
galaxies. Since this is the flux density regime we are probing with
our stacking analysis, one might hope to catch a glimpse of such
evolution. Indeed, the basic form of the evolution we see in \qfir\ is
consistent with the idea of \citeauthor{singal09}, and our spectral
index is consistent with that of the radio background ($\alpha^{\rm
8,000}_{\rm 22}\sim -0.6$, \citealt{fixsen09}).

\citet{swinbank08} used the {\sc galform} semi-analytical
galaxy-formation model to predict mild evolution of \qfir\ for
luminous starbursts due to the $\sim$5-Myr timelag between the onset
of star formation and the resulting supernovae, in keeping with the
mild evolution claimed by \citet{kovacs06} on the basis of 350-$\mu$m
observations of 15 SMGs. \citeauthor{kovacs06} suggested that a change
of the radio spectral index might account for the deviation in
$q$. Revisiting the {\sc galform} work of \citet{swinbank08} --
calculating \sfir\ and \qfir\ for galaxies with $L_{\rm
IR}>10^{12}$\,L$_{\odot}$ in a manner consistent with this work --
results in evolution of the form shown by the shaded area in
Fig.~\ref{stack} (lower panel), i.e.\ the evolution progresses in the
opposite sense to that seen, with quantitative agreement only in the
$2<z<3$ regime.

The evolution seen here has implications for all the areas that rely
on the FIR/radio correlation, e.g.\ the median redshift of SMGs
determined via the oft-used $S_{\rm 850\mu m}/S_{\rm 1,400MHz}$ flux
density ratio would be lower.  It will be interesting to discover
whether the evolution we report is confirmed by {\em Herschel} in the
deepest multi-frequency radio fields \citep[e.g.][]{om08}.

Taking the average properties of the 24-$\mu$m- and
250-$\mu$m-selected samples to be representative, so \qfir\ = $2.54
\pm 0.26$, we can suggest a simple transformation between radio flux
density and SFR, appropriate for samples in which radio-loud AGN
constitute only a small fraction: a radio flux density of $S_{\rm
1,400MHz}$ ($\rm W\,m^{-2}\,Hz^{-1}$) can be converted into a
bolometric IR flux (rest-frame 8--1,000\,$\mu$m in W\,m$^{-2}$) via:

\begin{equation}
 S_{\rm IR} = 10^{12.57 + (2.54\pm 0.26)} S_{\rm 1,400MHz} ,
\end{equation}

\noindent
and thence into a star-formation rate (SFR) via:

\begin{equation}
{\rm SFR} = \Psi \, 10^{-10} \, \frac{4\pi \, d_{\rm lum}^2 \, S_{\rm
IR}}{3.83 \times 10^{26}} \, {\rm M}_{\odot}\,{\rm yr}^{-1} ,
\end{equation}

\noindent
where $\Psi$ is 1.7 for a Salpeter IMF ($dN(m)/dm = -2.35$) covering
0.1--100\,M$_{\odot}$ \citep[][see also \citealt{sy83, tt86,
inoue00}]{condon92, kennicutt98a, kennicutt98b}.

\begin{figure}
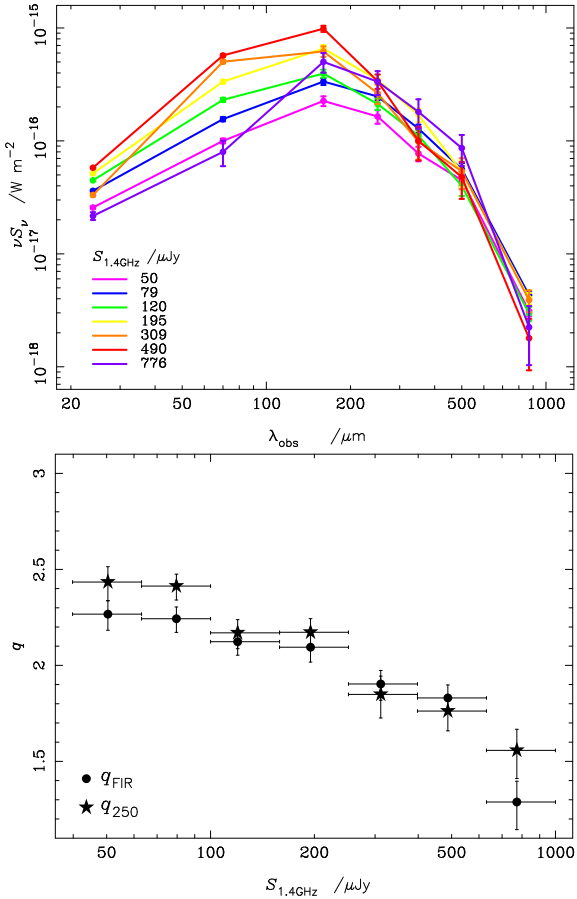

\begin{center}
  \includegraphics[scale=0.43,angle=270]{sed-radio.eps}
  \vspace*{0.5mm}

  \includegraphics[scale=0.43,angle=270]{radio-stack.eps}
  \caption{{\em Top:} $\nu_{\rm obs} S_{\nu_{\rm obs}}$ versus
    observed wavelength, determined by stacking into the IR/submm
    images at the positions of faint radio emitters, in seven flux
    density bins, spaced evenly in log$_{\rm 10} \, S_{\rm
      1,400MHz}$. Median flux densities in each bin are reported in
    the key. {\em Bottom:} \qtwofifty\ and \qfir\ in the seven flux
    density bins. $K$ corrections were not applied because complete
    redshift information was not available. The scale was chosen to
    match Fig.~\ref{stack} ({\em bottom}).}
  \label{radio-stack}
\end{center}
\end{figure}

\subsection{Stacking at the positions of sub-mJy radio galaxies}
\label{radstack}

The value of \qfir\ found in the previous section is weighted by the
types of galaxies found in the 24-$\mu$m catalogue. To investigate
whether \qfir\ is different for a radio-selected population, we stack
the FIR/submm images at the positions of 816 robust radio emitters
($>$5\,$\sigma$, $40 < S_{\rm 1,400MHz} < 1,000\,\mu$Jy), divided into
seven flux density bins, spaced evenly in log$_{\rm 10} \, S_{\rm
1,400MHz}$.

The FIR SEDs of the radio-selected galaxies are qualitatively similar
to those of low-redshift 24-$\mu$m-selected galaxies (compare the {\em
top} panels of Figs~\ref{stack} and \ref{radio-stack}). We find that
\qtwofifty\ and \qfir\ are lower for radio-selected galaxies than for
those selected in the mid-IR or FIR wavebands (Table~\ref{qtab}), as
one might expect given the selection criterion. However, the
difference is surprisingly small: for $40 < S_{\rm 1,400MHz} <
100\,\mu$Jy we obtain \qtwofifty\ = 2.42 and \qfir\ = 2.25, cf.\ the
equivalent (pre-$K$-correction) values for our FIR-selected galaxy
sample, \qtwofifty\ = $2.26\pm 0.35$ and \qfir\ = $2.40\pm 0.29$. Bear
in mind that, at a redshift of unity, \qfir\ would rise by $\sim$0.06
due to the radio $K$ correction and fall by $\sim$0.15 due to the
shift of $\lambda_{\rm rest}$ = 8--1000\,$\mu$m to lower observed
frequencies -- an overall shift of $\sim-0.1$. This suggests that star
formation plays a significant role in powering faint radio galaxies,
providing over half of their IR luminosity. Radio selection will
inevitably have favoured galaxies with recently injected relativistic
electrons so this is likely to be a lower limit. We find that
\qtwofifty\ and \qfir\ anti-correlate with radio flux density
(Fig.~\ref{radio-stack}, {\em lower} panel) and we suggest this is due
to the increasing prevalence of radio-loud AGN as one moves into the
mJy regime \citep[e.g.][]{ibar09}.

\section{Conclusions}
\label{conclusions}

We have defined a sample of 250-$\mu$m-selected galaxies using data
from BLAST. The noise is dominated by confusion, which severely
limits the number of robust detections and hampers the
identification of secure, unambiguous mid-IR or radio
counterparts. These problems were not always apparent when using an
objective, probabilistic approach to cross-identifying the galaxies.

We find that the most likely 24-$\mu$m counterparts to our small
sample of 250-$\mu$m galaxies have a median [interquartile] redshift
of 0.74 [0.25, 1.57].

At $z\approx 0.6$, where the BLAST 250-$\mu$m filter probes rest-frame
160-$\mu$m emission, we find no evidence for evolution of \qtwofifty\
relative to $q_{\rm 160}$ measured for the SINGS sample of local
galaxies.

We find that \sfir\ is better correlated with radio flux density than
$S_{\rm 250\mu m}$, and that $K$-correcting the radio luminosity using
measured spectral slopes reduces the scatter in the correlation.

\qfir\ -- the logarithmic ratio of bolometric IR and monochromatic
radio fluxes -- is determined for FIR-, mid-IR and radio-selected
galaxies. We provide a simple recipe to convert a radio flux density
into an instantaneous SFR, via bolometric IR luminosity, for mid-IR-
or FIR-selected samples that do not contain a large fraction of
radio-loud AGN.

\qfir\ is found to be similar for our 250-$\mu$m- and radio-selected
galaxies, which suggests that star formation is responsible for over
half of the IR luminosity in the latter, especially the faintest
radio galaxies ($S_{\rm 1,400MHz}<100\,\mu$Jy). This fraction could well
be higher given that radio selection favours galaxies with recent
injections of relativistic electrons.

Stacking into 610- and 1,400-MHz images at the positions of
24-$\mu$m-selected galaxies, we find no evidence that the spectral
slope at radio wavelengths evolves significantly across
$0<z<3$. However, we find tentative evidence that \qfir\ does evolve,
even across $0<z<1$ where incompleteness in the parent sample should
not be a serious issue. The evolution is of the form, \qfir\ $\propto
(1+z)^{-0.15\pm0.03}$, across the peak epoch of galaxy formation. This
has major implications for many techniques that rely on the FIR/radio
correlation.  We compare with semi-analytical model predictions and
speculate that we may be seeing an increase in radio activity amongst
ordinary, star-forming galaxies -- relative to their IR emission --
amongst that has been suggested may give rise to the radio background
\citep{fixsen09, singal09}.

It will remain difficult to associate galaxies detected in {\em
Herschel}/SPIRE surveys unambiguously with IR or radio counterparts,
since the modest increase in telescope aperture and the steep source
counts will ensure that confusion remains an issue
\citep{devlin09}. However, alongside 70-, 100- and 160-$\mu$m imaging
from {\em Herschel}/PACS and 450- and 850-$\mu$m SCUBA-2 data from the
15-m James Clerk Maxwell Telescope, it should be possible to fine-tune
the astrometry to $\sigma\sim 1$\,arcsec and to deblend sources in
deep SPIRE images. For the $\sim$1--2 deg$^2$ of FIR/submm coverage
planned for the fields with the deepest available radio and {\em
Spitzer} data, e.g.\ GOODS, the Subaru/{\em XMM-Newton} Deep Survey
and the Lockman Hole, with $\sigma_{\rm 1,400MHz} \ls
5$\,$\mu$Jy\,beam$^{-1}$ and $\sigma_{\rm 24\mu m} \ls 10$\,$\mu$Jy
(\citealt{biggs06, om08, miller08}; Morrison et al., in preparation;
Arumugam et al., in preparation), almost every deblended FIR source
should have relatively secure 24-$\mu$m and radio identifications, the
majority unambiguous, providing a complete redshift distribution and
less biased estimates of \qfir\ for FIR-selected galaxies.

\section*{Acknowledgements}

We thank John Peacock for his patient and good-natured assistance. We
acknowledge the support of the UK Science and Technology Facilities
Council (STFC), NASA through grant numbers NAG5-12785, NAG5-13301, and
NNGO-6GI11G, the NSF Office of Polar Programs, the Canadian Space
Agency, and the Natural Sciences and Engineering Research Council
(NSERC) of Canada.

\bibliographystyle{mnras}
\bibliography{refs}

\bsp

\end{document}